\documentclass[
superscriptaddress,
amsmath,amssymb, 
aps
]{revtex4-2}

\usepackage[T1]{fontenc} 
\usepackage{graphicx}
\usepackage{hyperref}
\usepackage[english]{babel} 
\usepackage[dvipsnames]{xcolor}
\hypersetup{
  colorlinks   = true, 
  urlcolor     = NavyBlue, 
  linkcolor    = OliveGreen,
  citecolor    = BrickRed  
}
\usepackage[mathlines]{lineno}
\usepackage{xcolor} 
\usepackage{gensymb} 
\usepackage{siunitx, booktabs} 
\DeclareSIUnit\angstrom{\text{\AA}}
\usepackage{multirow}
\makeatletter
\let\switch@array\relax
\makeatother

\begin{document}


\preprint{APS/tr-XRMR-Fe PRB}

\title{Depth-resolved magnetization dynamics in Fe thin films after ultrafast laser excitation}%

\author{Valentin Chardonnet}
\affiliation{Sorbonne Université, CNRS, Laboratoire de Chimie Physique – Matière et Rayonnement, LCPMR, 75005 Paris, France}

\author{Marcel Hennes}
 \affiliation{Sorbonne Université, CNRS, Institut des NanoSciences de Paris, INSP, F-75005 Paris, France}

\author{Romain Jarrier}
\affiliation{Sorbonne Université, CNRS, Laboratoire de Chimie Physique – Matière et Rayonnement, LCPMR, 75005 Paris, France}
 
\author{Renaud Delaunay}
\affiliation{Sorbonne Université, CNRS, Laboratoire de Chimie Physique – Matière et Rayonnement, LCPMR, 75005 Paris, France}

\author{Nicolas Jaouen}
\affiliation{Synchrotron SOLEIL, L'Orme des Merisiers Saint-Aubin, BP48, 91192 Gif-sur-Yvette, France}
 
\author{Marion Kuhlmann}
\affiliation{Deutsches Elektronen-Synchrotron, 22607 Hamburg, Germany}
 
\author{Cyril Leveillé}
\affiliation{Synchrotron SOLEIL, L'Orme des Merisiers Saint-Aubin, BP48, 91192 Gif-sur-Yvette, France}
\affiliation{Laboratoire Albert Fert, CNRS, Thales, Université Paris-Saclay, 91767 Palaiseau, France}

\author{Clemens von Korff Schmising}
\affiliation{Max Born Institut f\"ur Nichtlineare Optik und Kurzzeitspektroskopie, 12489 Berlin, Germany}

\author{Daniel Schick}
\affiliation{Max Born Institut f\"ur Nichtlineare Optik und Kurzzeitspektroskopie, 12489 Berlin, Germany}

\author{Kelvin Yao}
\affiliation{Max Born Institut f\"ur Nichtlineare Optik und Kurzzeitspektroskopie, 12489 Berlin, Germany}

\author{Xuan Liu}
\affiliation{Synchrotron SOLEIL, L'Orme des Merisiers Saint-Aubin, BP48, 91192 Gif-sur-Yvette, France}

\author{Gheorghe S. Chiuzbăian}
\affiliation{Sorbonne Université, CNRS, Laboratoire de Chimie Physique – Matière et Rayonnement, LCPMR, 75005 Paris, France}

\author{Jan Lüning}
\affiliation{Helmholtz-Zentrum Berlin für Materialien und Energie, 14109 Berlin, Germany}
\affiliation{Institute of Physics, Humboldt-Universität zu Berlin, Newtonstr. 15, D-12489 Berlin, Germany}

\author{Boris Vodungbo}
\affiliation{Sorbonne Université, CNRS, Laboratoire de Chimie Physique – Matière et Rayonnement, LCPMR, 75005 Paris, France}

\author{Emmanuelle Jal}
 \email{emmanuelle.jal@sorbonne-universite.fr}
\affiliation{Sorbonne Université, CNRS, Laboratoire de Chimie Physique – Matière et Rayonnement, LCPMR, 75005 Paris, France}

\date{\today}

\begin{abstract}
We performed time-resolved x-ray resonant magnetic reflectivity measurements on a laser-excited ferromagnetic Fe thin film to simultaneously probe the transient magnetic and structural depth profiles with nanometer spatial and femtosecond temporal resolution. Our results show that during the first picoseconds after optical excitation, the magnetization of the Fe layer is strongly inhomogeneous, especially in the vicinity of the buried interface. By comparing our experimental results to predictions based on the microscopic three-temperature model and simulations of laser-induced spin-currents, we demonstrate that local and non-local angular momentum transfer phenomena take place simultaneously. After a few picoseconds, the magnetization relaxes back to equilibrium while the total thin film thickness starts oscillating periodically, with a maximum dilation of approximately $1.3\%$ of the entire thin film thickness due to laser-induced stresses.

\end{abstract}

\maketitle

\section{\label{sec:level1}Introduction}
The observation of an ultrafast laser-induced demagnetization in Ni thin films by Beaurepaire \textit{et al.,} in 1996~\cite{beaurepaire1996ultrafast} laid the foundation for the field of femtomagnetism, which aims at understanding how ultrashort light pulses impact the magnetic properties of materials \cite{kirilyuk_ultrafast_2010,malvestuto_ultrafast_2018,scheid_light_2022}. This scientifically intriguing phenomenon has a high potential to spur disruptive technological development. During the last two decades, a variety of models were developed to explain the ultrafast quenching of magnetization in metallic thin films after femtosecond optical excitation \cite{koopmans_explaining_2010, battiato_superdiffusive_2010, zhang_laser-induced_2018, scheid_light_2022}.
They can be classified into two broad categories, relying either on a local or a non-local transfer of angular momentum \cite{zhang_laser-induced_2018}. In the case of a simple \textit{3d} magnetic layer sandwiched between two non-magnetic-metallic thin films, local angular momentum transfer will induce a depth-dependent demagnetization that is mainly proportional to the exponential absorption profile of the laser intensity, yielding a stronger loss of magnetization close to the top interface \cite{wieczorek_separation_2015,you_revealing_2018}. In contrast, non-local effects result from spin currents injected into the adjacent layers, generating an additional demagnetization close to both interfaces \cite{battiato_superdiffusive_2010,battiato_theory_2012,eschenlohr_ultrafast_2013,wieczorek_separation_2015}. Thus, assessing the transient magnetic depth profile with femtosecond temporal and nanometer spatial resolution would allow disentangling both mechanisms. So far, this has been an experimental challenge. Studies that tried to distinguish between local and non-local phenomena relied on classical time-resolved magneto-optical experiments in specific geometries that allow probing the top and bottom interfaces separately \cite{hofherr_speed_2017,shokeen_spin_2017}, employed non-linear optical effects such as time-resolved second harmonic generation \cite{melnikov_ultrafast_2011,wieczorek_separation_2015,chen_competing_2019,elliott_transient_2022}, or used element specific techniques \cite{turgut_controlling_2013, hennes_element-selective_2022, schmising_ultrafast_2023}. To date, none of these experiments have provided a direct measurement of the transient magnetic profile inside a single \textit{3d} ferromagnetic layer. 

X-ray resonant magnetic reflectivity (XRMR) is a well-established technique for analyzing static magnetic profiles in thin films \cite{kravtsov_complementary_2009, jal_magnetization_2013, macke_magnetic_2014, jal_interface_2015}. This synchrotron-based approach couples x-ray reflectivity with x-ray magnetic circular dichroism (XMCD) to measure reflected intensities that depend on the charge density and magnetization. 
With the advent of x-ray free electron lasers (XFELs) \cite{malvestuto_ultrafast_2018, abela_perspective_2017,huang_features_2021, kovalchuk_european_2022} and more powerful high-harmonic generation (HHG) sources \cite{HHG_source_2021}, it is now possible to extend such experiments to the femtosecond time domain. One advantage of tr-XRMR compared to  tr-XMCD is that magnetic sensitivity can also be retrieved when using linearly polarized light in the scattering plane \cite{jal_magnetization_2013,jal_reflectivite_2013}. While this experimental transverse configuration is very similar to classical time-resolved transverse magneto-optical Kerr effect experiments (tr-T-MOKE), which are usually performed at a fixed angle and energy, tr-XRMR relies on data gathered over a broad angular or energy range, which allows one to retrieve the full magnetic and structural depth profiles \cite{macke_magnetic_2014}. Indeed, tr-T-MOKE measurements performed in the extreme ultraviolet range (XUV, 40\,eV to 200\,eV) paved the way to element-specific analysis of ultrafast magnetization dynamics in alloys \cite{La-O-Vorakiat_ultrafast_2009,mathias_probing_2012,turgut_controlling_2013,hofherr_ultrafast_2020,vaskivskyi_element-specific_2021, jana_atom-specific_2023,schmising_ultrafast_2023,von_korff_schmising_direct_2024,moller_verification_2024}, but are not directly proportional to the magnetization \cite{zusin_direct_2018,jana_analysis_2020,Richter_Relationship_2024,probst2024unraveling}, and do not allow for the retrieval of the magnetic depth profile \cite{gutt_probing_2017}. 
Hennecke \textit{et al.} recently overcame these limitations by using a continuous probing spectrum to quantitatively resolve transient magnetic depth profiles during all-optical helicity-independent switching \cite{hennecke_transient_2025}, thereby demonstrating the capability of tr-XRMR for depth-resolved studies. Nevertheless, XUV radiation limits the available probing depth, hampering studies of deeply buried interfaces. To benefit from a larger penetration depth and better element selectivity, one can use shorter wavelengths, such as soft x-ray (200\,eV to 2000\,eV), by extending tr-XRMR to the  $L_{2,3}$ edges of $3d$ metals.  

So far, tr-XRMR studies in the soft x-ray range have been mainly performed at the femtoslicing beamline of BESSY II \cite{thielemann-kuhn_ultrafast_2017, golias_ultrafast_2021, awsaf_element-selective_2024, gordes_accelerated_2025}, where the angular range is limited because of the low available photon flux, which prevents the retrieval of transient depth magnetic profiles \cite{jal_structural_2017}.
Recently, we have shown that a tr-XRMR experiment at the Fe $L_{3}$ edge and for specular reflection angles up to $30\degree$ is possible at XFEL, using the third order of FLASH2 \cite{chardonnet_toward_2021}.
Since tuning the accelerator to scan different x-ray energies can be difficult and time-consuming, we performed angular scans at a fixed energy, as is done in static synchrotron experiments \cite{jal_magnetization_2013, jal_interface_2015}. In this paper, we demonstrate how quantitatively analyzing our tr-XRMR angular scans at the $L_{3}$ edge enables us to track simultaneously the depth-dependent magnetic and structural dynamics of an optically excited 18\,nm thick Fe layer with subnanometer spatial resolution.

\section{Experimental Details}
\begin{figure}[t!]
\includegraphics[width=7cm]{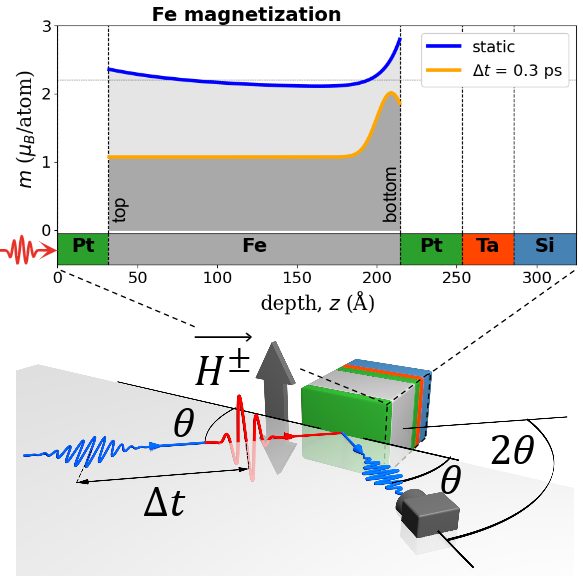}
\caption{\label{fig1} Schematic of the time-resolved magnetic reflectivity (tr-XRMR) experiment and derived magnetic profile. The x-ray probe pulses with linear polarization are shown in blue, and the pump pulse in red. Both impinge on the sample at an incidence angle $\theta$. A transverse magnetic field is applied perpendicularly to the scattering plane. 
The top panel shows the derived magnetic profile for the unpumped sample (blue) and 300 fs after the excitation (orange).}
\end{figure}

Our experiments were carried out on the FL24 FLASH2 beamline, using a home-built reflectometer \cite{chardonnet_toward_2021}, on a sample prepared by DC-sputtering, with nominal layer thicknesses of Si/Ta(3\,nm)/Pt(3\,nm)/Fe(15\,nm)/Pt(3\,nm) \footnote{A static characterization of the samples by XRMR was performed at the SEXTANTS beamline of SOLEIL, yielding the following multilayer structure: Si/Ta(3.3\,nm)/Pt(3.2\,nm)/Fe(18.4\,nm)/Pt(3.9\,nm), in good agreement with the nominal composition.}. For different time delays, $\Delta t$, between the optical pump and the x-ray probe pulses, we measured the intensities of the specular reflection $I^+(\theta,\Delta t)$ and $I^-(\theta,\Delta t)$ as a function of the reflection angle $\theta$ for a positive and negative magnetic field applied in the transverse direction, respectively (Fig. \ref{fig1}). 
To separate structural and magnetic information, one can analyze the average reflected intensity $R=\frac{I^+ + I^-}{2}$ which is mainly proportional to electronic and structural quantities such as the density, thickness and roughness, or evaluate the asymmetry $A=\frac{I^+-I^-}{I^+ + I^-}$, which is proportional to the ratio of magnetic to charge contributions, and from which magnetic information can be retrieved \cite{kortright_resonant_2013,macke_magnetic_2014,kumberg_ultrafast_2023}. 

To excite the sample, we used linearly polarized infrared (IR) pump pulses of 800\,nm with a duration of 50\,fs, and a normal incidence footprint of 230\,\textmu m at full width half maximum. These pulses were collinear with the soft x-ray probe pulses (Fig. \ref{fig1}). Since a variation of the scattering angle, $\theta$, will inevitably change the laser excitation fluence, we adjusted the incoming intensity and restricted the angular range of our time-resolved measurement to a grazing reflection angle of  $\theta \in [20\degree,35\degree]$, leading to maximal deviations around a mean absorbed fluence of $2\pm0.4$\,mJ/cm$^2$ (see \cite{chardonnet_toward_2021} for more details). To probe the sample, we used linearly-in-plane-polarized photons having an energy of 701\,eV with a 7\,eV bandwidth, resonant to the Fe $L_3$ edge, with a duration of 80\,fs, and a normal incidence footprint of 80\,\textmu m at full width half maximum. The energy broadening resolution results from the use of the third harmonic generated by the FEL FLASH2 \cite{chardonnet_toward_2021}. We used the typical FLASH2 pulse structure with trains of pulses every 100\,ms. We chose an intra-train frequency of 200\,kHz to have 40 x-ray pulses spaced by \qty{5}{\micro\second} in each train. These were then combined with 20 IR pulses at a frequency of 100\,kHz so only every second x-ray pulse is associated to an IR pulse \footnote{We were sharing the beamtime with another experiment that needed 40 IR pulses per train, so for practical reasons, we also had 40 pulses per train on the sample but probed only the first 20 IR pulses. It was not affecting our sample temperature and experimental results}. This allows measurement of pumped and unpumped reflected intensities for both magnetic field directions, $I^\pm_{\text{p}}$ and $I^\pm_{\text{up}}$. Note that this $I^\pm_{\text{up}}$ implies probing the sample approximately 5\,\textmu s after the last IR pump, where we assume that the sample is back to its equilibrium static state. These $I^\pm_{\text{p}}(\theta,\Delta t)$ and $I^\pm_{\text{up}}(\theta,\Delta t)$, were measured as a function of the reflection angle $\theta \in [21\degree,34\degree]$  for nine different time delays $\Delta t$ between the pump and the probe (-1, 0, 0.08, 0.15, 0.30, 0.50,  1, 7.3 and 13.5\,ps); from which we extracted pumped and unpumped reflectivity and asymmetry signals ($R_{\text{p}}(\theta,\Delta t)$, $R_{\text{up}}(\theta,\Delta t)$ and $A_{\text{p}}(\theta,\Delta t)$, $A_{\text{up}}(\theta,\Delta t)$ respectively). All raw experimental data and extracted signals are presented in detail in the Supplemental Material \cite{SupMat}.

\begin{figure*}[]
\includegraphics[width=18cm]{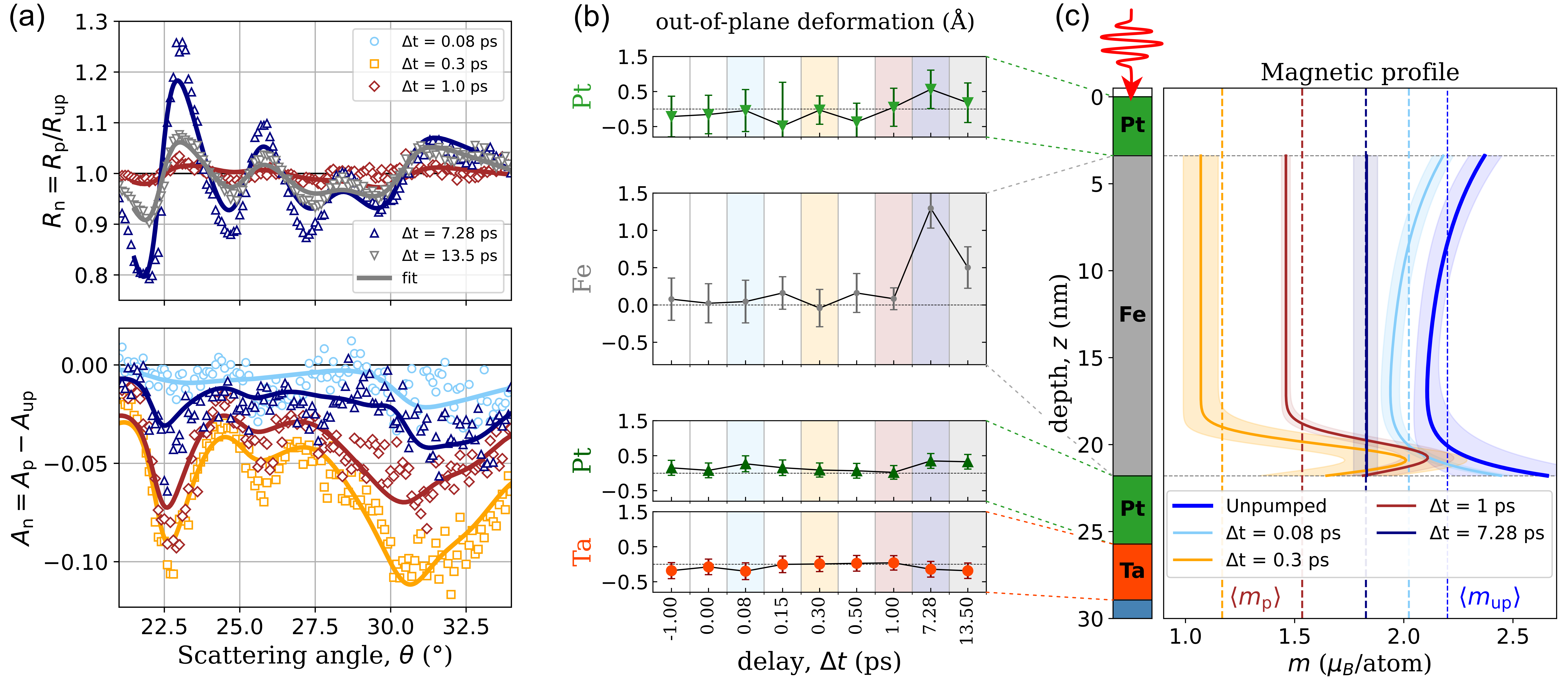}
\caption{\label{fig3} Fitting results with respective out-of-plane deformation and magnetic depth profile for different time delays. (a) Experimental (points) normalized reflectivity, $R_\text{n}$ (top panel) and normalized magnetic asymmetry $A_\text{n}$ (bottom panel) for different delays (see legend) with the corresponding best fit (lines) (b) Derived out-of-plane deformation (pumped thickness minus unpumped thickness) in \r{A} for the different layers of the sample as a function of each delay $\Delta t$. The background colour represents the different delays $\Delta t$ plotted in (a). (c) Magnetic moments $m$ of the Fe layer are shown on the horizontal axis as a function of the sample depth, $z$, on the vertical axis, where, on the left, the sample structure is shown. Derived depth magnetic profiles for 4 different delays, with the unpumped profile equal to that of Fig. \ref{fig1}. The dotted vertical lines represent the mean of the magnetization, $\langle m_\text{p} \rangle$ and $\langle m_\text{up} \rangle$, through the entire Fe layer.}
\end{figure*}

\section{Results and quantitative analysis}

By fitting $I^\pm$ simultaneously as a function of the reflection angle, photon energy, and x-ray polarization, one can eventually derive the structural and magnetic depth profiles \cite{jal_magnetization_2013,macke_magnetic_2014}. 
We used the Dyna package, a matrix formalism that calculates reflectivity using a classical description based on Maxwell's equations and a permittivity derived from the quantum scattering amplitude \cite{elzo_x-ray_2012}. For the resonant scattering factors of our different layers, we used tabulated data for the Ta, Pt and Si charge scattering factors \cite{henke_x-ray_1993,cxro_atomic}, and data from C. T. Chen work \cite{chen_experimental_1995} for the Fe charge and magnetic resonant scattering factors \cite{SupMat}. While the fits are performed simultaneously on $I^\pm$, we display the fitting results on the reflectivity $R$ and magnetic asymmetry $A$ to separate the electronic and structural information from the magnetic one.
As explained in our previous work \cite{jal_interface_2015} and in the Supplemental Material \cite{SupMat}, the magnetic film can be subdivided into several layers to retrieve depth-dependent magnetic information. To reduce the number of free parameters, we describe the magnetization amplitude by an analytical function depending on the film depth $z$ \cite{hennecke_ultrafast_2022}. Motivated by previous studies showing a steep enhancement of the Fe magnetization close to the top and the bottom interfaces \cite{jal_magnetization_2013, jal_interface_2015}, we used a function with an exponential enhancement at each interface  \cite{SupMat}, which we will refer to as \emph{two-exponential} function in the following, to fit the static measured intensities of the reflected beam $I^\pm_{\text{up}}$.
The derived magnetic profile is shown in blue in Fig. \ref{fig1} and \ref{fig3}~c. Note that the agreement between the experimental and simulated curves is excellent, as shown in the Supplemental Material \cite{SupMat}. Furthermore, this depth-resolved magnetic profile has been extracted by fitting XRMR experimental data obtained at the synchrotron SOLEIL for several photon energies around the Fe $L_3$ edge and also allows us to reproduce experimental data obtained for another XRMR experimental geometry using circularly polarized x-rays and a longitudinal applied magnetic field \cite{SupMat}, corroborating the choice of our derived magnetic profile.

To analyze the time-dependent averaged reflectivity and magnetic asymmetry, $R_\text{p}$ and $A_\text{p}$, we used the same matrix formalism and procedures. For each time delay, we fitted $I^\pm_\text{p}$ by taking the static parameters as initial guesses. We obtained satisfactory results by fitting only the thickness of each layer and the Fe-layer magnetization depth profile. Among the numerous magnetic models tested for fitting purposes \cite{SupMat}, we focused on four different magnetization profile functions. In addition to a \emph{1-layer} and \emph{two-exponential} magnetization profile used for unpumped data, we defined one \emph{single-exponential} function, which mimics the local loss of magnetization due to the exponential absorption profile of the laser intensity; and a \emph{single-Gauss} function, which aims to capture the additional non-local demagnetization at the bottom interface, induced by superdiffusive spin-currents. Only the function yielding the best agreement for each delay was eventually taken into consideration to describe the temporal evolution of the system \cite{SupMat}.

Figure \ref{fig3}~a shows the experimental normalized reflectivity signal $R_\text{n} = R_\text{p} / R_\text{up}$ (top panel) and the normalized asymmetry $A_\text{n} = A_\text{p} - A_\text{up}$ (bottom panel) for different time delays $\Delta t$ (points) as well as the best corresponding fits (line). We observe a good agreement between the experimental data and the fit for both $R_\text{n}$ and $A_\text{n}$. The corresponding fitting parameters, relative thicknesses, and magnetization profiles are presented in Figs. \ref{fig3}~b and \ref{fig3}~c, respectively.  
The difference between the extracted pumped and unpumped thicknesses, which corresponds to the compression or dilatation of each layer (out-of-plane deformation), is plotted for each layer as a function of $\Delta t$ in Fig. \ref{fig3}~b. Our results show that from a structural perspective, the thin film remains unaffected up to 1\,ps. Then, at $\Delta t = 7.3$\,ps, we observed an increase in thickness of the Fe layer by $1.3 \pm \SI{0.3}{\angstrom}$, the top Pt layer by $0.4 \pm \SI{0.5}{\angstrom}$, and the bottom Pt layer by $0.4 \pm \SI{0.2}{\angstrom}$. For longer time delays, a long relaxation back to equilibrium takes place, as can be observed for $\Delta t= 13.5$ ps.  
This layer-resolved information can be integrated over the entire thin film thickness and is plotted in Fig. \ref{fig4} as a function of the time delay (red dots).

Figure \ref{fig3}~c shows the magnetic profiles extracted from the fits as a function of the sample depth. Before and just after the laser excitation, the sample displays a \emph{two-exponential} magnetic profile (blue for unpumped and light blue for $\Delta t = 80$ fs). Before the pump, the average moment of 2.20\,$\mu_B$ corresponds to the bulk value, while 80 fs after the pump pulse, the average moment decreases to 2.02 $\mu_B$. The notable increase at the Pt interfaces is due to the symmetry breaking happening at the interface \cite{ohnishi_surface_1983,lu_first-principles_2013}. During the first hundreds of fs, the profile changes markedly, it is reduced homogeneously over the first \qty{15}{\nano\meter}, and presents a \qty{3}{\nano\meter} wide peak close to the bottom interface (yellow and dark red, $\Delta t$= 300 fs and 1 ps). Finally, on picosecond timescales, the fitting results in a completely homogeneous magnetic profile (dark blue, $\Delta t$= 7.3 ps), with a reduced average moment of 1.8 $\mu_B$. This simpler \emph{1-layer} model has been chosen to describe our data because the more complex fit functions do not improve the fit quality, considering our signal-to-noise ratio.
The average ultrafast demagnetization of the Fe layer $\langle m_\text{p} \rangle$  can be extracted by averaging the depth profiles and are shown by dashed line in Fig. \ref{fig3}~c. When plotted as a function of the delay, we retrieve a demagnetization curve (Fig. \ref{fig5}~a) that is consistent with earlier measurements that lack depth sensitivity \cite{koopmans_explaining_2010, jal_structural_2017}. The fitting by a double exponential function gives a demagnetization time of 180 fs \cite{SupMat}, in agreement with previous work \cite{koopmans_explaining_2010, jal_structural_2017}. Our extracted transient depth magnetic profiles can be interpolated to eventually obtain a 2D map of the normalized magnetization $m_\text{p}/m_\text{up}$ as a function of depth and $\Delta t$ (Fig. \ref{fig5}~b).

\section{Discussion}

\begin{figure}
\includegraphics[width=7.5cm]{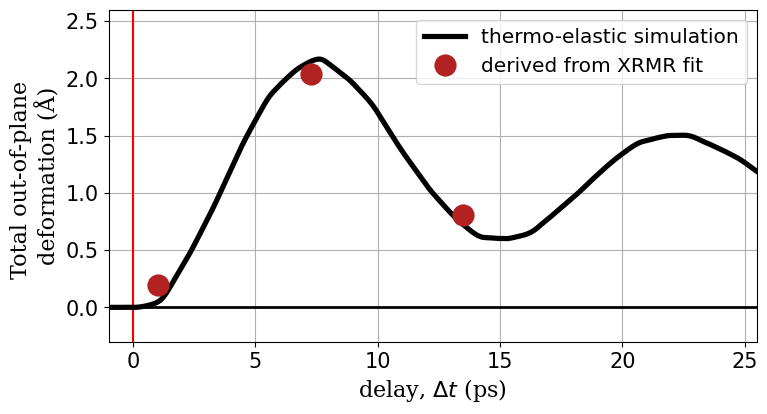}
\caption{\label{fig4} Laser-induced total out-of-plane deformation as a function of time delay, extracted from the $R_n$ fits of Fig. \ref{fig3}~a, in red points. The black line is the total deformation calculated using the udkm1Dsim toolbox. }
\end{figure}

\begin{figure}
\includegraphics[width=8cm]{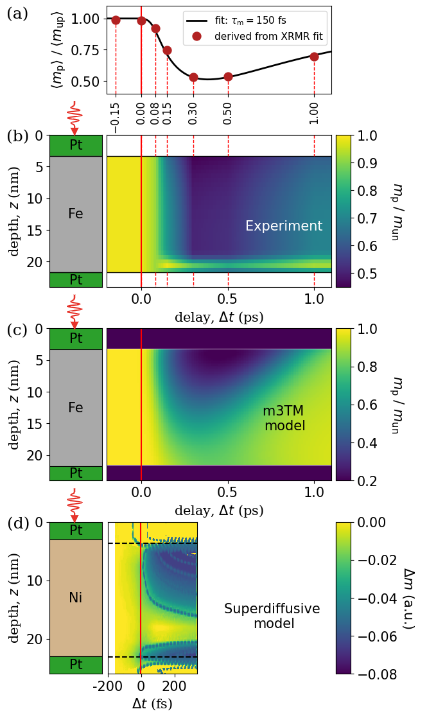}
\caption{\label{fig5} Extracted laser-induced magnetization profile and calculations based on the m3TM and superdiffusive simulations. (a) Normalized mean magnetization,  $\langle m_\text{p} \rangle/ \langle m_\text{up}\rangle$, as a function of time delay (red points), with a double exponential fit (black line \cite{SupMat}). (b) 2D-map of the normalized magnetization $m_\text{p}/m_\text{up}$, extracted from the fit of our experimental data (Fig. \ref{fig3}~a), as a function of the depth, $z$ (vertical axis) and time delay $\Delta t$ (horizontal axis). The red vertical dotted lines indicate the measured $\Delta t$. (c) 2D-map of the normalized magnetization $m_\text{p}/m_\text{up}$ for the m3TM model, calculated for our sample structure using the udkm1Dsim toolbox \cite{schick_udkm1dsim_2024}. (d) 2D-map of the relative magnetization $\Delta m = m_\text{p}-m_\text{up}$ for the superdiffusive spin model in a 15 nm thick Ni sample, adapted from \cite{eschenlohr_ultrafast_2013} with the permission of the authors.}
\end{figure}

This quantitative analysis of our tr-XRMR data demonstrates that structural information (Fig. \ref{fig4}) and transient magnetic profiles (Fig.\ref{fig5}~b) can be retrieved simultaneously. 
It confirms the two different time scales observed in our previous work \cite{chardonnet_toward_2021}. Structural modifications set in after approximately 1\,ps, whereas ultrafast demagnetization effects are much faster and exhibit a strongly inhomogeneous profile along the thin-film depth.

Our total out-of-plane deformation of \qty{2}{\angstrom} observed after \qty{7.3}{\pico\second} corresponds to a maximum of the normalized reflectivity time traces shown in our previous article (Fig. 6a of \cite{chardonnet_toward_2021} and Fig. S11 of the supplemental material \cite{SupMat}), which displays a damped oscillating behavior with a period around 14.6 ps. This experimental periodicity is in very good agreement with the time, $\tau=$\qty{14.7}{\pico\second}, needed for an acoustic longitudinal strain wave to travel back and forth within all our thin layers at the speed of sound \cite{time_speedSound}. 
To simulate this total out-of-plane deformation due to the optically induced thermal strain, we used the udkm1Dsim toolbox \cite{schick_udkm1dsim_2021, schick_udkm1dsim_2024, mattern_concepts_2023}.
The simulation results are plotted as a black line in Fig. \ref{fig4} \cite{SupMat}. The agreement with the experimental data is very good for both oscillation periodicity and amplitude. This confirms the strength of tr-XRMR for probing thickness modification due to laser-induced stresses. 
Note that while the strain waves start immediately at the interface \cite{SupMat}, we can't access the transient strain depth profile in reflectivity. We can resolve only the total thickness change over one elemental layer.

In contrast to the structural changes, happening on ps timescales, Fig. \ref{fig5}~a and b show a large demagnetization effect of 50 $\%$ between 300 and 500 fs that is almost homogeneous in the upper \qty{15}{\nano\meter} of the Fe layer. Surprisingly, in the last \qty{3}{\nano\meter} close to the bottom Pt interface, the loss of magnetization is much smaller and gives rise to a persistent magnetization band, followed by a decrease of the magnetization in the very close vicinity (< 1 nm) of the Pt layer. As shown in the Supplemental Material \cite{SupMat}, this \emph{persistent magnetization band} feature is essential to reproduce accurately the experimental unpumped asymmetry $A_{up}$. Furthermore, this decrease of magnetization observed at the Fe/Pt bottom interface agrees with studies showing stronger and faster demagnetization of the induced Pt moment compared to the transition metal magnetic moment \cite{schmising_ultrafast_2023,yamamoto_ultrafast_2019,vaskivskyi_element-specific_2021,richter_spectroscopic_2025}. 
Note that even though we took care to have the same Pt layer at the top and bottom Fe interfaces, those interfaces can not be considered equal since they have  (i) different roughness (\qty{8}{\angstrom} and \qty{3}{\angstrom} for the top and bottom interfaces resp.) and (ii) different excitation profiles reaching those two interfaces \cite{SupMat}. This explains why the magnetic profiles at both Pt interfaces are not identical.

To gain a better understanding of the underlying microscopic mechanisms, we have extended our simulations using the udkm1Dsim toolbox \cite{schick_udkm1dsim_2024} by implementing equations of the local microscopic 3 temperature model (m3TM) \cite{koopmans_explaining_2010} into the heat diffusion calculation. As shown in Fig. \ref{fig5}~c, the results of this transient magnetization depth profile simulation do not agree with our observations. While the time scale of the ultrafast demagnetization is correctly assessed for the first 10 nm inside the Fe layer, the magnetization recovery is much faster, and there is no bottom interface effect \cite{SupMat}. 
Even though the m3TM has allowed for reproducing experimental data obtained by optical MOKE \cite{koopmans_explaining_2010,roth_temperature_2012} the observed differences between our experimental results and this local m3TM could be due to (i) heat diffusion \cite{wieczorek_separation_2015,kuiper_nonlocal_2011,steinbach_exploring_2024}, which has been poorly studied, or (ii) other mechanisms that are not taken into account such as spin-orbit coupling \cite{zhang_laser-induced_2000,tows_many-body_2015}, dynamic exchange splitting reduction \cite{mueller_feedback_2013,kraus_ultrafast_2009}, or electron and lattice dynamics \cite{zahn_intrinsic_2022}. Note, however, that none of the aforementioned models can explain the strongly inhomogeneous magnetization profile observed at the bottom interface.

As highlighted in the introduction, an additional loss of magnetization at the bottom interface is predicted by models that rely on non-local angular momentum transfer mechanisms. Indeed, transient magnetic profiles have been calculated for the superdiffusive model in Ni thin films on top of an Al layer \cite{battiato_superdiffusive_2010} or sandwiched between Pt layers \cite{eschenlohr_ultrafast_2013}, and for Fe thin films on top of an Al layer \cite{battiato_theory_2012}. Irrespective of the magnetic material and adjacent metallic layer that were scrutinized, a demagnetization in the vicinity of the bottom interface is observed, which is due to the majority spin electrons traveling into the adjacent non-magnetic metallic layer. This gives rise to a \emph{persistent magnetization band} inside the magnetic layer. Qualitatively, this is exactly what we are seeing in the last nanometers close to the bottom Pt layer (Fig. \ref{fig5}~b). However, from a quantitative perspective, our transient magnetic profiles differ from theoretical predictions, as becomes apparent when comparing our experimental results to the simulations of Eschenlohr et al.  \cite{eschenlohr_ultrafast_2013} (Fig. \ref{fig5}~d). In this superdiffusive simulation, the demagnetization near the surface is inhomogeneous and relaxes within the first 10 nm, whereas we observe homogeneous, flat demagnetization up to the first 15 nm in our Fe layer. Furthermore, the \emph{persistent magnetization band} appearing in the transient magnetization profile near the interface is much sharper in our case, roughly 1.5 nm wide, with a decrease in magnetization at the bottom interface observed only over the last nanometer. These discrepancies could be due to differences in the mean free path of Ni and Fe minority electrons (2\,nm and 0.5\,nm, respectively) while the majority spin electrons both travel ballistically around 7\,nm before they scatter \cite{zhukov_lifetimes_2006}. The poor crystallinity of our sample as well as differences in the interface roughnesses might also impact the results, as shown by Lu \textit{et al.,} \cite{lu_interface_2020}. From this study, it seems that the large roughness we found at the top interface could be one reason why we are not seeing any hint of spin-polarized electron current at this top interface. Furthermore, the optical excitation profile (Fig. S11.b) results in much stronger local absorption at the top interface, thereby enhancing local effects.

Clearly, the m3TM and superdiffusive models presented in Fig. \ref{fig5}~c and d, provide an oversimplified picture of the mechanisms at work in our laser-excited thin film. 
Considering other ultrafast magnetization microscopic explanations also allows to explain only different parts of our experimental transient depth profile. If one takes only ballistic transport into account, a recent study \cite{ashok_signatures_2025} demonstrates that the change in magnetization is homogeneous in depth. While this would explain our homogeneous flat demagnetization in the first 15 nm of the sample, it does not account for interface effects. We could also consider our \emph{persistent magnetization band} to result from a spin accumulation at a ferromagnetic to metallic interface that counteracts the demagnetization process, as has been predicted through a simple $s-d$ model \cite{beens_s-d_2020}. All the mechanisms discussed so far are mainly Stoner-like models, but Heisenberg-like mechanisms could also partly explain our depth demagnetization profile. The coexistence of an optically induced paramagnetic phase, close to the surface where the excitation is stronger, with the initial ferromagnetic phase could explain the slow magnetization recovery \cite{eich_band_2017, you_revealing_2018}. Furthermore, the effect of optically-induced collective magnon excitation \cite{schmidt_ultrafast_2010,carpene_ultrafast_2015, zusin_direct_2018, gort_early_2018, dusabirane_interplay_2024, jana_fluence_2025, weissenhofer_ultrafast_2025, zheng_ultrafast_2025, rouzegar_femtosecond_2025} on depth magnetic profiles has not been studied so far, but would probably smooth the magnetization profile on picosecond timescales.

\section{Conclusion}
Our study demonstrates that magnetic and structural depth profiles can be assessed simultaneously following optical excitation, combining sub-nanometer spatial and sub-picosecond temporal resolution. As detailed in this paper, our experiment provides evidence for an optically induced strain wave in the thin film, consistent with thermo-elastic simulations. 
Magnetic effects occur on shorter time scales and set in long before lattice expansion effects can be observed. 
The extracted in-depth transient magnetization is strongly inhomogeneous during the first picoseconds after optical excitation. We find a flat demagnetization profile that expands up to 15 nm and a spin accumulation with a typical width of a few nanometers located close to the bottom Pt interface. Comparing these results to existing theoretical models shows that both local and non-local angular momentum transfer are at work during demagnetization, and that no single model allows for reproducing our observations on their own. This highlights the need to combine several microscopic theories to unravel the mechanisms behind ultrafast demagnetization. 

Compared to recent tr-XRMR studies performed in the XUV range \cite{hennecke_ultrafast_2022, hennecke_transient_2025}, our work shows that using soft x-rays allows for simultaneously probing structural changes and subtle variations of the magnetic profile close to the buried interface of thicker films.
This paves the way for a better understanding of the interplay between lattice and magnetic degrees of freedom on ultrafast timescales \cite{reid_beyond_2018,von_reppert_spin_2020, dornes_ultrafast_2019,pfaff_strain_2025}.
Finally, our work highlights the strength of tr-XRMR to directly quantify spin transport phenomena inside magnetic thin films, which, so far, have mostly been detected indirectly by probing ultrathin multilayers in an element-selective-fashion \cite{hennes_element-selective_2022, geneaux_spin_2024}. In contrast, tr-XRMR has the potential to yield the full transient depth profile, even within single-element layers, which can provide valuable guidance for interface and multilayer design in future opto-spintronic devices.

\begin{acknowledgments}
The research leading to these results was supported by the project CALIPSOplus under Grant Agreement No. 730872 from the EU Framework Program for Research and Innovation HORIZON 2020 and was carried out at the FL24 beamline of FLASH2 at DESY, a member of the Helmholtz Association (HGF). Access to Synchrotron SOLEIL and beamline SEXTANTS through proposal ID 20160880 for the characterization of static properties of the Fe thin film is acknowledged. The authors acknowledge the use of the MagnetronMOKE@LCPMR facility supported by Sorbonne Université and the CNRS for the sample growth. The authors are grateful for the financial support received from CNRS GotoXFEL, MOMENTUM and MEDYNA ANR-20-CE42-0012-01. V.C. acknowledges the financial support from the Doctoral School ED 388 Chimie-Physique et Chimie Analytique de Paris Centre. C.v.K.S. acknowledges financial support from the Deutsche Forschungsgemeinschaft (DFG, German Research Foundation), Project ID No. 328545488, TRR 227, and Project No. A02. D.S. acknowledges financial support by the Leibniz Association through the Leibniz Junior Research Group Grant No. J134/2022 .

\end{acknowledgments}


\clearpage


\setcounter{page}{1}
\setcounter{section}{0}
\setcounter{figure}{0}
\setcounter{table}{0}
\setcounter{equation}{0}
\renewcommand{\thefigure}{S\arabic{figure}}
\renewcommand{\thetable}{S\arabic{table}}
\renewcommand{\theequation}{S\arabic{equation}}
\renewcommand{\figurename}{Figure}
\renewcommand{\tablename}{Table}

\vspace*{2cm}
{\centering\Large\textbf{SUPPLEMENTAL MATERIAL}\par}
\vspace*{1cm}


In this supplemental material, we give details about:
\begin{enumerate}
    \item Basic knowledge on x-ray resonant magnetic reflectivity (section \ref{sec:basXRMR}. XRMR)
    \item The fitting procedure employed to extract magnetic depth profiles from static XRMR measurements performed at the beamline Sextants (SOLEIL). See section \ref{sec:SOLEIL_fit}.
    \item The fitting procedure employed to derive meaningful magnetic profiles using the time-resolved XRMR data gathered at the free-electron laser FLASH2, as well as key parameters described in the main article, such as $R_\text{p}$, $R_\text{up}$, $A_\text{p}$ and $A_\text{up}$. See section \ref{sec:FLAS_fit_300fs}.
    \item The optical laser excitation profile of the Fe sample used in this study. See section \ref{sec:Optic_exc_prof}.
    \item The parameters used to simulate the strain map and microscopic 3 temperature model (m3TM) using to the udkm1Dsim toolbox 
    \cite{schick_udkm1dsim_2021, schick_udkm1dsim_2024}. See section \ref{sec:udkm1Dsim}.
    
\end{enumerate}
Additional information can be found in \cite{chardonnet_etude_2022, chardonnet_toward_2021}.

\section{\label{sec:basXRMR}Fundamentals of XRMR}
To guide readers unfamiliar with XRMR, we summarize the concepts, and notions needed to fully grasp the results of this work. First, we will review the pros and cons of the different experimental geometries used in this work. Then we will explain to what extent the reflectivity, $R$,  and asymmetry, $A$, discussed in this work, are proportional to the charge and magnetic contributions. Finally, we will specify the framework of our matrix formalism used to fit our experimental data. For more information on XRMR, we point the reader to those reviews \cite{kortright_resonant_2013,macke_magnetic_2014} as well as our previous work \cite{elzo_x-ray_2012, jal_reflectivite_2013, chardonnet_etude_2022, chardonnet_toward_2021}. \\

XRMR is a combination of x-ray reflectivity and x-ray magnetic circular dichroism, which gives two different reflected intensities, $I^\pm$, depending on the interaction of the incoming electromagnetic x-ray wave with the magnetization of the studied thin film. To measure those intensities, several experimental configurations can be used \cite{jal_reflectivite_2013}. In this work, we mainly use the transverse configuration, which consists of measuring the reflected intensity of an X-ray beam linearly polarized in the scattering plane for two opposite directions of a magnetic field applied in the sample plane and perpendicular to the scattering plane (see Fig. 1b of the article). In this configuration, without any polarization analysis of the reflected beam, we are only sensitive to the transverse component of the magnetization. This implies that we can't distinguish a decrease in magnetization from a rotation of the magnetization away from the transverse axis.

For the static characterization of our sample, we have used an additional experimental configuration with circularly polarized light and a longitudinal applied field (in the sample and scattering planes). In this longitudinal-circular geometry, we are sensitive to magnetization in 3D, meaning that, through a matrix-formalism analysis, we can differentiate a decrease in magnetization from a rotation of magnetization away from its longitudinal direction. \\

From the resonant atomic scattering factor derived by Hannon \cite{hannon_x-ray_1988} (equation \ref{eq:Hannon}) and in the kinematical limit \cite{kortright_resonant_2013}, the reflected intensities $I^\pm$ can be approximated to equation \ref{eq:Is}.
\begin{equation}
    f_{res}(E,\theta)=\big( \vec{e_i}^{*} \cdot \vec{e_f} \big)F_{c,r}  - i\cdot \vec{m}\cdot \big( \vec{e_i}^{*} \wedge \vec{e_f} \big)F_{m,r}
    \label{eq:Hannon}
\end{equation}
\begin{equation}
\begin{aligned}
    I^\pm(E,\theta) &\propto |f^{res}(E,\theta) |^2 \\
                    &\propto F^*_{c,r}F_{c,r}\pm 2m \mathrm{Re}\{F^*_{c,r}F_{m,r}\}+m^2F^*_{m,r}F_{m,r} \\
                    &\propto C^2 \pm CM+M^2 
    \label{eq:Is}
\end{aligned}
\end{equation}
where, $E$ is the energy, $\theta$ is the angle between the wavector and the sample plane,  $\vec{e_{i}}$ and $\vec{e_{f}}$ are the unit polarization vectors of incident and reflected wavevectors, $F_{c,r}$ and $F_{m,r}$ are the complex charge and magnetic resonant scattering factors, and $\vec{m}$ is a unit vector along the axis of local magnetization. From equation \ref{eq:Is} and \ref{eq:R-A}, one directly sees that the reflectivity $R$ defined in the article is mainly sensitive to the square of the charge term since even at resonance the magnetic term squared will be negligible. The asymmetry ratio $A$ is then sensitive to the magnetic term normalized by the charge term.
\begin{equation}
\begin{aligned}
    R &= \frac{I^+ + I^-}{2} = \frac{C^2+M^2}{2}\propto C^2 \\
    A &= \frac{I^+ - I^-}{I^+ + I^-} \propto \frac{M}{C}
    \label{eq:R-A}
\end{aligned}
\end{equation}
While we used $R$ and $A$ to separate electronic and structural information from magnetic information, since they are not strictly proportional to charge and magnetization, respectively, we fitted both intensities $I^\pm$ simultaneously. This ensures that when structural modifications occur, they are accounted for in the asymmetry signal $A$, and vice versa; if a strong change in magnetization occurs, it is accounted for in the reflectivity $R$. \\

To fit our experimental reflected intensities $I^\pm$, we used the Dyna package, a matrix formalism that calculates reflectivity using a classical description inspired from the Parratt formalism \cite{parratt_surface_1954} and based on Maxwell’s equations and a permittivity tensor derived from the quantum scattering amplitude that can be linked to the resonant atomic scattering factor derived by Hanon \cite{hannon_x-ray_1988} and shown in equation \ref{eq:Hannon}. This formalism accounts for all polarization channels and a 3D magnetic vector $\vec{m}$ \cite{elzo_x-ray_2012}. 
To simulate the magnetic reflected intensities $I^\pm$, one needs to implement the charge and magnetic scattering factors $F_{c,r}$ and $F_{m,r}$ that both have a real and imaginary part that is related to the dispersive and absorptive part of the refractive index, as described in \cite{elzo_x-ray_2012,macke_magnetic_2014}. As said in the article, we used tabulated data for the charge scattering factors of Ta, Pt, and Si \cite{henke_x-ray_1993,cxro_atomic}, while data from C. T. Chen work \cite{chen_experimental_1995} were used for the $F_{c,r}$ and $F_{m,r}$ resonant scattering factors of Fe. 
To adjust the simulated intensities to the experimental ones, the free parameters are the thickness, roughness, densities, and the 3D magnetization vector $\vec{m}$. This implies that we assume the scattering factors' properties do not change upon laser illumination. Furthermore, as stated above, in the transverse experimental configuration used for our time-resolved study, we are sensitive only to changes in the transverse magnetization amplitude $m$. In other words, we scale the atomic magnetic resonant scattering factor $F_{m,r}$ by a factor $m$ for each magnetic layer to fit our reflected intensities.

\section{\label{sec:SOLEIL_fit}Static sample characterization}

Before measuring the optically induced ultrafast magnetization dynamics of our Fe thin film, we characterized our sample in static conditions by performing XRMR at the SEXTANTS beamline of SOLEIL. We measured XRMR for several photon energies around the Fe $L_3$ edge (704.7, 705.1, 706.1, 707.1, 708.1, and 709.1\,eV) with an energy resolution of 140 meV, and for the same transverse geometry as the one described in the main text.  
Compared with our previous analysis \cite{chardonnet_toward_2021}, the magnetic profile has been further refined for the present study.

\begin{figure}[h!]
\includegraphics[width=11cm]{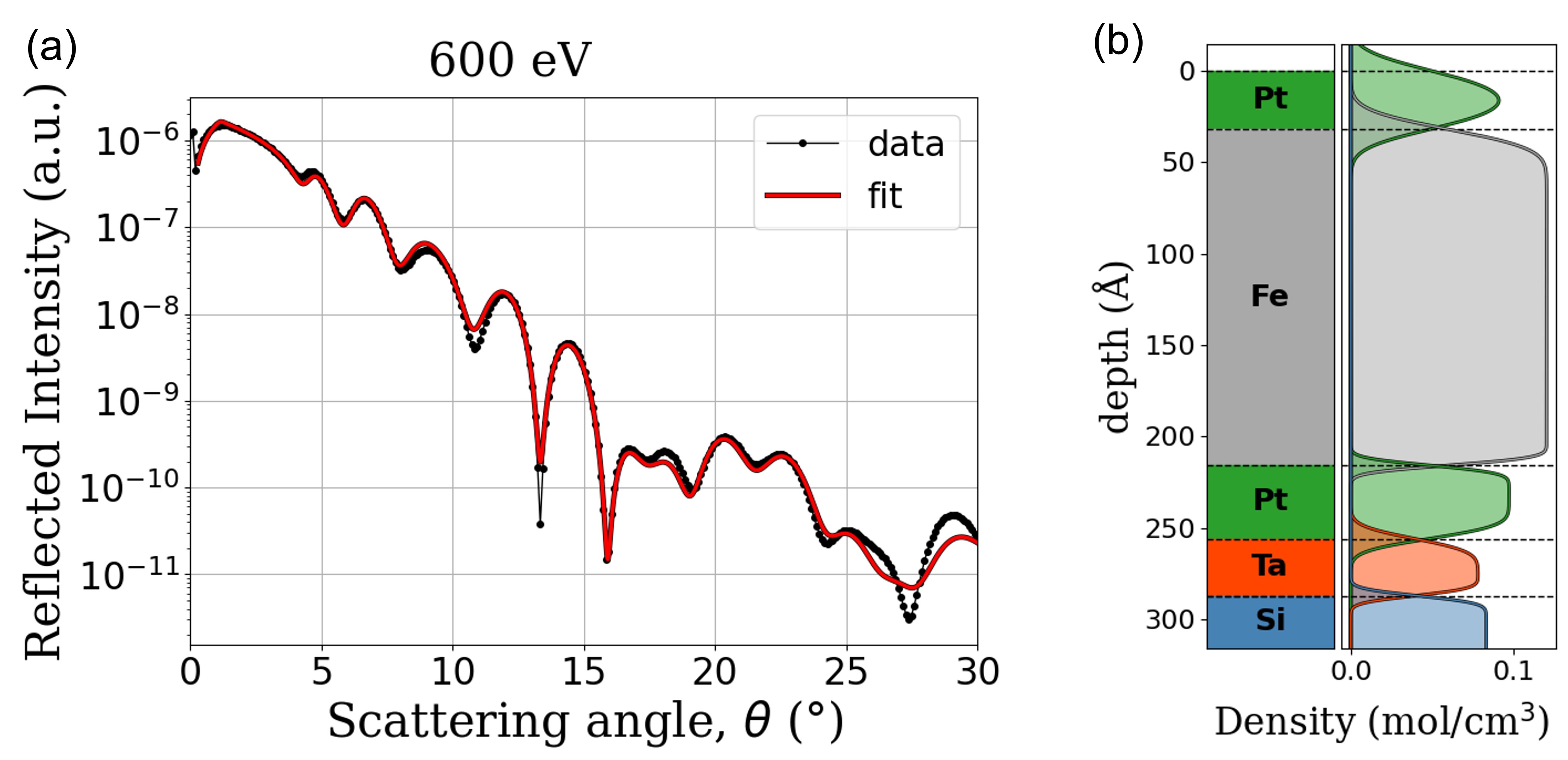}
\caption{\label{fig:struct} (a) Static reflected intensity measured at SOLEIL for an x-ray energy of 600 eV (black circles) and best fit (red line) corresponding to the structural profile shown in (b): schematic representing the layer densities as a function of the sample depth $z$. }
\end{figure}

\begin{table}[h]
\begin{center}
\begin{tabular}{p{1cm}p{3cm}p{3cm}p{3cm}}
    \hline\hline
     & Density         & Thickness & Roughness     \\
     & (mol.cm$^{-3}$) & (nm)      & (nm)          \\
    \hline
    Pt & $0.097$ & $3.2 \pm 0.2$ & $0.9 \pm 0.1$ \\
    Fe & $0.120$ & $18.4\pm 0.2$ & $0.8 \pm 0.2$ \\
    Pt & $0.097$ & $3.9 \pm 0.2$ & $0.3 \pm 0.1$ \\
    Ta & $0.078$ & $3.3 \pm 0.2$ & $0.5 \pm 0.2$ \\
    Si & $0.083$ &                 & $0.3 \pm 0.2$ \\
    \hline\hline
\end{tabular}
\caption{\small{Structural parameters derived from the fit of off-resonance reflectivity data shown in Fig.  \ref{fig:struct}~a and used for the schematics of Fig. \ref{fig:struct}~b.}}
\label{tab:struct}
\end{center}
\end{table}

To retrieve structural and magnetic information, the experimental signals were fitted using a well-established matrix formalism explained briefly in the \hyperref[sec:basXRMR]{Fundamentals of XRMR} section and in more details in \cite{elzo_x-ray_2012}.
While we are simultaneously fitting $I^+$ and $I^-$, for clarity, we show only the fitting results for the asymmetry ($ A = \frac{I^+- I^-}{I^+ + I^-}$). 
Using the structural parameters presented in \cite{chardonnet_toward_2021}, Fig. \ref{fig:struct}, and Table \ref{tab:struct}, we began fitting the data using a homogeneous depth profile, i.e., a single magnetic layer with constant magnetization.
As seen in Fig. \ref{fig : Static fit SOLEIL-1}, for all x-ray energies, there is a discrepancy between the experimental magnetic asymmetry $A$ (black dots) and the fit results for this \emph{1-layer} magnetic model (red line), especially between $23\degree$ and $28\degree$.

\begin{figure}[b!]
\includegraphics[width=17cm]{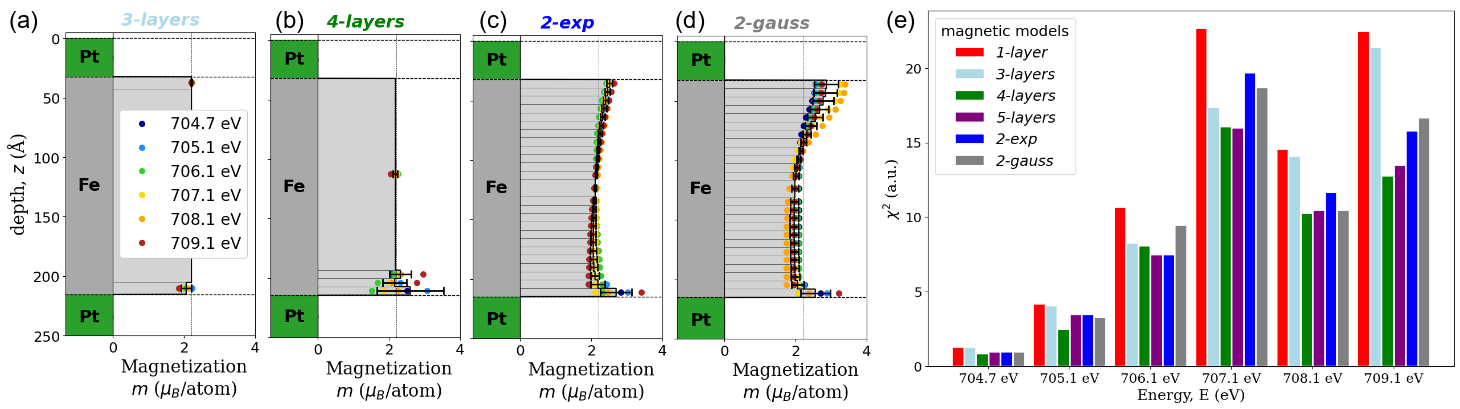}
\caption{\label{fig : static magnetic profile}Magnetic moment of the Fe layer as a function of the sample depth $z$ for different magnetic models: (a) \emph{3-layers}, (b) \emph{4-layers}, (c) \emph{two-exponential}, (d) \emph{two-gauss} for all the probing x-ray energies (colored dots). Black lines and error bars represent the average value over all those energies. (e) displays the value of the minimisation function $\chi^2$ for the different x-ray energies and models discussed in the text.}
\end{figure}

To improve agreement between the experimental and calculated asymmetries, we divided the Fe layer into several magnetic layers.  In a first attempt, we separated the Fe layer into 3 magnetic layers, consisting of one thick, central bulk layer and two thin interface layers \cite{jal_interface_2015}. Since magnetic interface effects are supposed to take place in the first four monolayers $\approx$6\r{A} \cite{ohnishi_surface_1983,lu_first-principles_2013} which is smaller than the roughness of the interface (8\r{A}), we have chosen a fixed interface layer thickness of 7\r{A}. The magnetic profiles derived from the best fits corresponding to this \emph{3-layers} model are shown in Fig. \ref{fig : static magnetic profile}~a  for all probing energies. This figure shows the magnetic moment of the Fe layer on the horizontal axis as a function of depth on the vertical axis. The sample structure with the different magnetic layers is shown on the left. Close to the top Pt interface, the magnetization changes slightly, whereas near the bottom Pt interface, there are substantial changes. The agreement between the fit and the experiment is slightly improved as summarized in Fig. \ref{fig : static magnetic profile}~e. 
To see if another model can better reproduce our experimental data, we then divided our Fe layer into 4 magnetic layers as shown in Fig. \ref{fig : static magnetic profile}~b. The best corresponding fits for all the x-rays probing energies are shown with green lines in Fig. \ref{fig : Static fit SOLEIL-1} and the agreement quality is shown in Fig. \ref{fig : static magnetic profile}~e. By looking at the extracted layer-dependent magnetization profile for the different x-ray energies (Fig. \ref{fig : static magnetic profile}~b), one can see that, on average, this \emph{4-layers} model gives an increase of the magnetization at the bottom interface.  

\begin{figure}[t!]
\includegraphics[width=15.5cm]{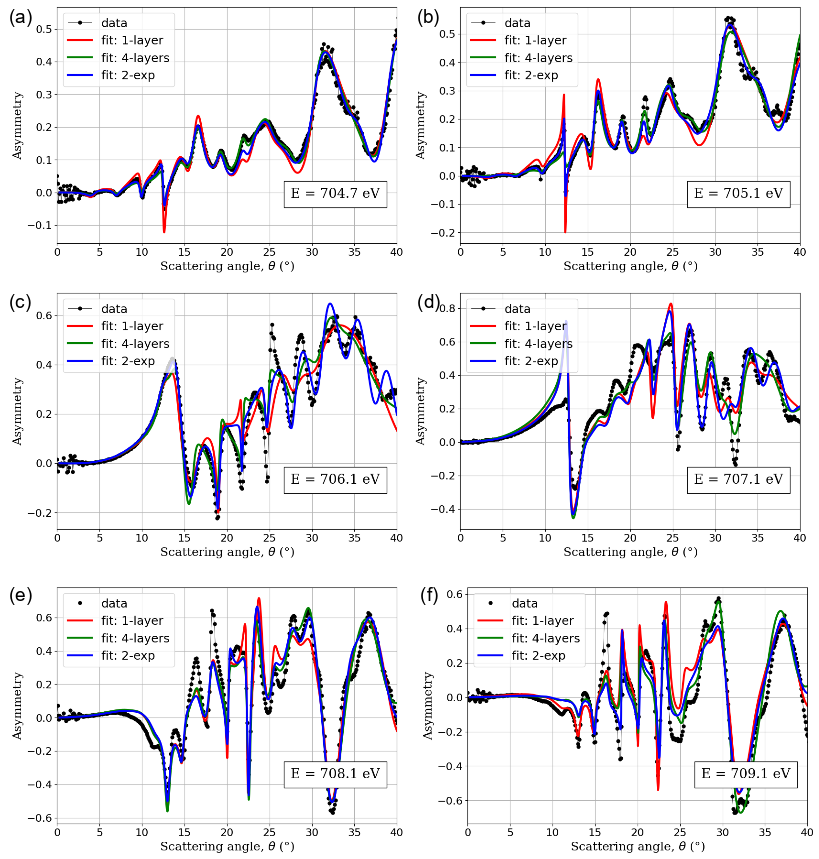}
\caption{\label{fig : Static fit SOLEIL-1} Static asymmetry measured at SOLEIL (black circles) with an energy resolution of 140 meV and best fits for different magnetic models: \emph{1-layer} i.e. \emph{homogeneous} (red line), \emph{4-layers} (green line) and \emph{two-exponential} (blue line) for several photon energies around the Fe $L_3$ edge of (a) 704.7, (b) 705.1, (c) 706.1, (d) 707.1, (e) 708.1 and (f) 709.1 eV. }
\end{figure}

As shown in Fig. \ref{fig : static magnetic profile}~e, if we continue to divide our Fe layer into more sub-magnetic layers, the agreement between the best-fit and the experimental data does not improve. Inspired by other XRMR works \cite{macke_magnetic_2014, hennecke_ultrafast_2022}, we therefore decided to test depth-dependent magnetic profile functions. This allowed us to approach a continuous magnetic profile with a reasonable number of free parameters. Since the matrix formalism requires a certain discretization of the Fe layer into sub-layers and to keep the interface symmetric, we therefore split our Fe layers into 12 layers of 7\r{A}, 1 middle layer of $\approx$16\r{A} and 12 layers of  7\r{A} as shown in Fig. \ref{fig : static magnetic profile}~c and d. The thickness of the middle layer is a fitting parameter while the thicknesses of 7\r{A} layers are fixed. Since we expect changes mainly at the interfaces, we tried two depth magnetization functions: a \emph{two-exponential} one, $m_\text{2-exp}$, and a \emph{two-Gauss} one, $m_\text{2-gauss}$: 

\begin{align}
m_{2-exp}(z) &= m_\text{up} \cdot e^{-\frac{z-z_\text{up}}{\sigma_\text{up}}} + m_{0} + m_\text{low} \cdot e^{\frac{z-z_\text{low}}{\sigma_\text{low}}} 
\label{equa:d-exp} \\
m_{2-gauss}(z) &= m_\text{up} \cdot e^{-\Big(\frac{z-z_\text{up}}{\sigma_\text{up}}\Big)^2} + m_{0} + m_\text{low} \cdot e^{-\Big(\frac{z-z_\text{low}}{\sigma_\text{low}}\Big)^2}
\label{equa:d-gauss}
\end{align}
with $m_{0}$, the bulk magnetization, $m_\text{up}$ ($m_\text{low}$) and $\sigma_\text{up}$ ($\sigma_\text{low}$), the amplitude and width of the magnetization increase close to the upper (lower) interfaces that are fixed at a depth derived from the structural fit ($z_\text{up}=3.2$\,nm and $z_\text{low}=21.6$\,nm) \cite{chardonnet_toward_2021}.
Figure \ref{fig : static magnetic profile}~d and e display the magnetic profile corresponding to the best fit for each probing energy for the \emph{two-exponential} and \emph{two-gauss} function, respectively. While the minimization function is on average similar for both \emph{two-exponential} and \emph{two-gauss} models, the average profile of the \emph{two-gauss} shows larger discrepancies between the energies, and the large change of magnetization at the top interface is in contradiction with our previous simpler model showing no change of magnetization (Fig. \ref{fig : static magnetic profile}~a and b). The model showing the best agreement with our experimental data, with the smallest error bars in the derived averaged magnetic profile, is therefore the \emph{two-exponential} for which the corresponding fits are shown as blue lines in Fig. \ref{fig : Static fit SOLEIL-1}. An average of all the parameter values found for the different energies gives the profile shown in Fig. 1 and 2~c of the main article (blue line), which follows the equation \ref{equa: double exp soleil} with error bars coming from the standard deviation of the parameters found for each x-ray energy (indicated in parentheses).

\begin{align}
    m_{2-exp}(z) &= 0.30(\pm0.05) \cdot e^{-\frac{z-3.2}{6.0(\pm1.0)}} + 2.07(\pm0.03) + 0.70(\pm0.10) \cdot e^{\frac{z-21.6}{1.1(\pm0.5)}} 
    \label{equa: double exp soleil}
\end{align}

Table \ref{tab:static-model} summarizes the different models used and details: the magnetic layer structure, the number of magnetic fitting parameters, the colors of the corresponding calculated asymmetry in Fig. \ref{fig : Static fit SOLEIL-1}, corresponding magnetic depth profile in Fig. \ref{fig : static magnetic profile}, and the value of the minimisation function $\chi^2_{ave}$ averaged over all x-ray energies. 

\begin{table}[h]
\begin{center}
\begin{minipage}{0.95\textwidth}
\small{
\begin{tabular}{p{2.5cm}p{5.5cm}p{1.9cm}p{2.5cm}p{2.3cm}p{1.5cm}}
Magnetic model & magnetic layers (\SI{}{\nano\meter}) & $\#$ parameters  & colors in fig. \ref{fig : Static fit SOLEIL-1} & depth profile & $\chi^2_{ave}$ \\ \hline
\emph{1-layer} & ~~~~~~~~~~~~~~~~~~~~$x$ & 1 & red & not shown & 12.7\\
\emph{3-layers} &  0.7 |~~~~~~~~~~~~$x$~~~~~~~~~~~~| 0.7  & 3 & not shown & \ref{fig : static magnetic profile}~a & 11.1\\
\emph{4-layers} & ~~~~~~~~~~~~~~~~~~~~$x$ | 0.7 | 0.7 | 0.7 & 4 & green & \ref{fig : static magnetic profile}~b & 8.5 \\
\emph{5-layers} &  0.7 | 0.7 | ~~$x$~~ | 0.7 | 0.7 & 5 & not shown & not shown & 8.7\\
\emph{two-exponential} &  $(12 \times |0.7|)$ | $x$ | $(12 \times |0.7|)$ & 5 & blue &\ref{fig : static magnetic profile}~c & 9.9\\
\emph{two-gauss}&  $(12 \times |0.7|)$ | $x$ | $(12 \times |0.7|)$ & 5 & not shown &\ref{fig : static magnetic profile}~d & 10.0\\
\end{tabular}}
\caption{\small{Summary of the different magnetic profile models tested, with the sub-layer cut-outs used, the number of free parameters for the fit, the colors of the best fit in fig. \ref{fig : static magnetic profile}, the corresponding magnetic profiles resulting from these models, and the value of the minimisation function $\chi^2_{ave}$ averaged over all x-ray energies.}}
\label{tab:static-model}
\end{minipage}
\end{center}
\end{table}

\begin{figure}[b!]
\includegraphics[width=13
cm]{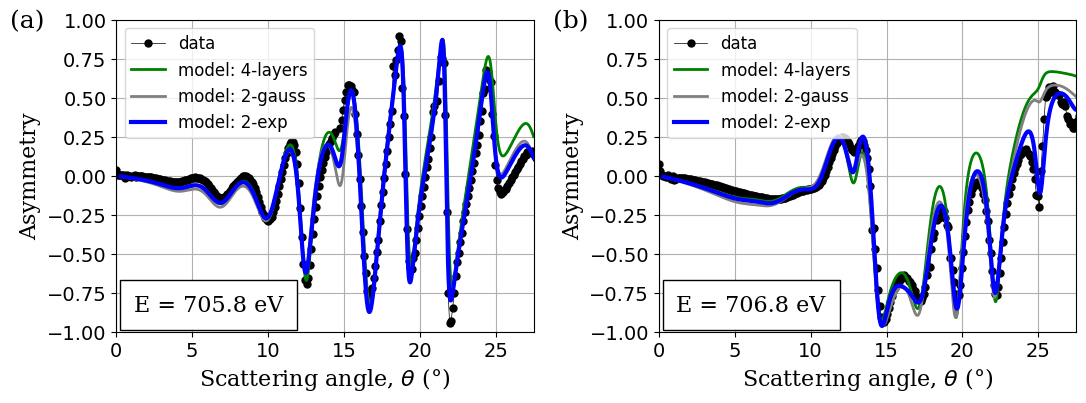}
\caption{\label{fig : test CL}Data and models for a longitudinal configuration at \SI{705.8}{\electronvolt} (a) and \SI{706.8}{\electronvolt} (b), i.e. using circular x-ray polarization and a longitudinally applied field (in the scattering and sample plane). Three magnetic models are shown: the \emph{4-layers} one (green line), the \emph{two-exponential} one (red line), and the \emph{two-gauss} one (grey line)}
\end{figure}

To validate our \emph{two-exponential} model, we calculated the asymmetry with the magnetization profile of equation \ref{equa: double exp soleil} for a different experimental configuration using circularly polarized light with a longitudinal applied field (in the sample plane and scattering plane) for two soft x-ray energies: 705.8 and 706.8 eV. As shown in figure \ref{fig : test CL}, the agreement with the experimental data (black dots) is very good. Note that this good agreement is not coming from a fit but directly from the model of equation \ref{equa: double exp soleil} (blue line). Furthermore, we observe that the \emph{two-exponential} simulation clearly outperforms the other ones (\emph{4-layers} in green and \emph{two-gauss} in grey) in terms of agreement with the experimental data. One advantage of an experimental configuration with circularly polarized light and a longitudinal field is its sensitivity to magnetization rotation out of the plane \cite{jal_noncollinearity_2015}. The good agreement with our experimental data for a model that changes only the magnetization amplitude therefore indicates that there is no out-of-plane magnetization rotation at the Pt interfaces.

\begin{figure}[h!]
\includegraphics[width=17
cm]{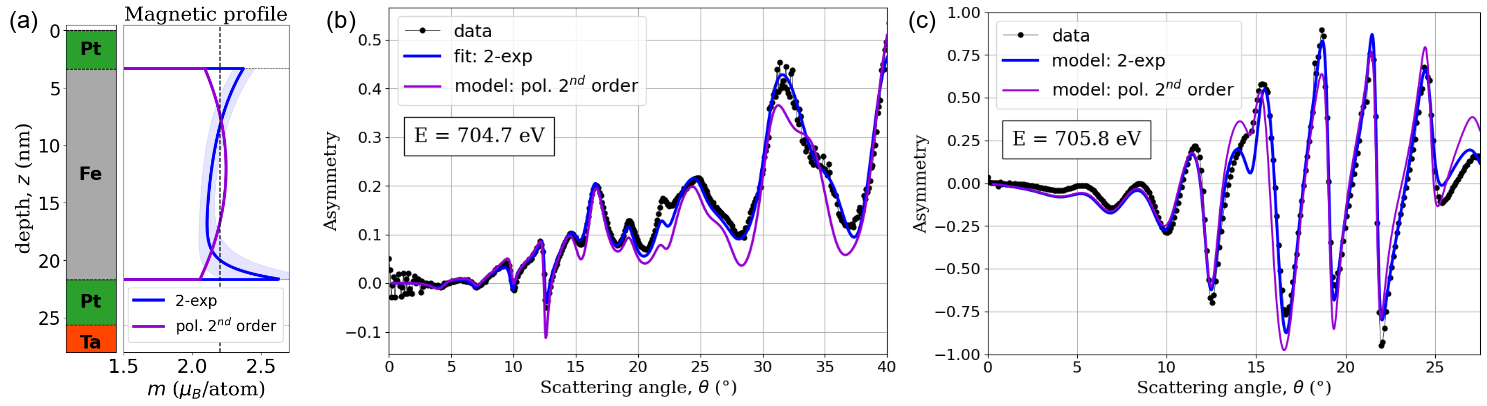}
\caption{\label{fig:test_polynom} (a) Magnetic profile for the \emph{two-exponential} (blue line) and 2$^\text{nd}$ order polynomial model (violet line). (b) Data and models for a transverse configuration at \SI{704.7}{\electronvolt}. (c) Data and models for longitudinal configuration at \SI{705.8}{\electronvolt}.}
\end{figure}

Finally, note that in a recent study on ferrimagnetic samples \cite{hennecke_ultrafast_2022, hennecke_transient_2025}, the static magnetic depth profile was fit to a 2nd-order polynomial, which shows a decrease in magnetization at both interfaces. We tested such a model, and as shown in fig. \ref{fig:test_polynom}, the agreement between the data and the fit is worse than in the case of our \emph{tow-exponential} model.

\section{\label{sec:FLAS_fit_300fs}Time-resolved data and fitting procedure}

\begin{figure}[b!]
\includegraphics[width=10cm]{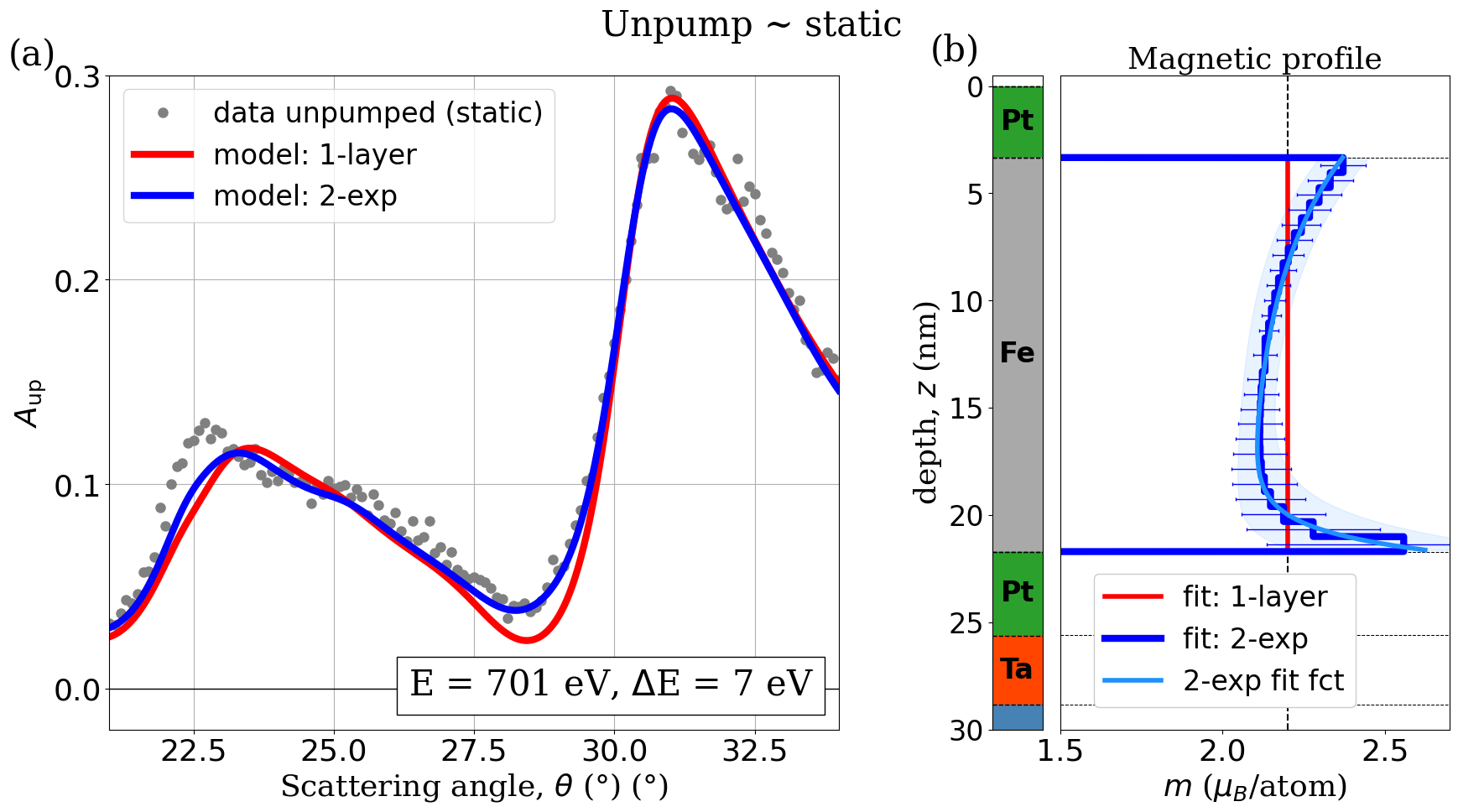}
\caption{\label{figS6} (a) Static asymmetry $A_\text{up}$ measured at FLASH (grey circles) using x-rays tuned to 701 eV with an energy resolution of 7\,eV. The calculated asymmetry for the \emph{1-layer} model is shown as red line, the \emph{two-exponential} model already used in the previous section \ref{sec:SOLEIL_fit}as a blue line. (b) Magnetic depth profile corresponding to the model reproducing the unpumped (static) data shown in (a) for the \emph{1-layer} (red) and \emph{two-exponential} (blue) simulations. In light blue, the \emph{two-exponential} function without the sub-layer cut as expressed in equation \ref{equa: double exp soleil}.}
\end{figure}

Static measurements were also performed at FLASH. In addition to confirming that meaningful XRMR data can be gathered on this FEL using energies at the Fe $L_3$  edge, these measurements also allowed us to check the initial sample state between each pump pulse, ($\approx5\,\mu$s after the pump). Those static measurements will be termed "unpumped", $A_\text{up}$, in the following and are plotted in fig. \ref{figS6}~a (black dots). The calculated asymmetries for a homogeneous profile, i.e, for the \emph{1-layer} model (red line), and for the \emph{two-exponential} model derived from SOLEIL analysis (blue line) are also displayed. As expected, the double exponential model shows a better agreement with the experimental data, despite the different probing energy and energy resolution (E = 701 $\pm$ 7 eV for FLASH). The corresponding profiles are plotted in Fig. \ref{figS6}~b, where we also show the \emph{two-exponential} function corresponding to equation \ref{equa: double exp soleil}. For greater clarity in Figures 1 and 2 of the article, we show only the function without the sub-layer cut. Note that to simulate the 7\,eV energy resolution, we calculated the reflectivity curves for 5 different energies (697, 699, 701,703, 705\,eV) and multiplied them by a Gaussian profile centered at 701\,eV and having a full-width at half maximum of 7 eV \cite{chardonnet_etude_2022}.\\

To analyse the pumped reflectivity $R_\text{p}$ and asymmetry $A_\text{p}$ data, we used the same matrix formalism and procedure as for the static data presented above. As explained in the main article, for each time delay, we fitted only the thickness of each layer, using the static parameters as starting guesses, and tried different depth magnetic profiles. First, we used different numbers of magnetic layers, from 1 to 5, as in the previous section \ref{sec:SOLEIL_fit} (not shown). For time delays between 100\,fs and 1\,ps, the model that best agreed with experimental data was the \emph{4-layer} one, which showed a persistent magnetization in the 2nd layer from the bottom upon laser excitation. Even though the decrease of magnetization at the bottom interface was only observed in the last 7\r{A}, this result seems similar to simulations predicting superdiffusive spin currents at the interface \cite{battiato_superdiffusive_2010, battiato_theory_2012,eschenlohr_ultrafast_2013}. Therefore, when switching to a more continuous profile using magnetic functions, we tested a \emph{single-Gauss} function $m_{gauss}$ (equation \ref{equa:gauss}), which aims to capture non-local demagnetization at the bottom interface. To also test the local loss of magnetization due to the exponential absorption profile of the laser intensity, we defined
one \emph{single-exponential} function $m_{exp}$, equation \ref{equa:surf-exp}. Table \ref{tab: modele_dynamique} summarizes the fit functions discussed in the article and hereafter.

\begin{table*}[h!]
\begin{ruledtabular}
\begin{tabular*}{\textwidth}{
  @{\extracolsep{\fill}}
  l 
  >{$\displaystyle}r<{\vphantom{\sum_{1}{N}}$} 
  @{\extracolsep{0pt}}
  >{$\displaystyle{}}l<{\vphantom{\sum_{1}{N}}$} 
  @{\extracolsep{\fill}}
  >{\refstepcounter{equation}(\theequation)}r
  @{}}
Magnetic profile types & \multicolumn{2}{l}{Magnetic profile equations} & \multicolumn{1}{c}{} \\
\midrule
\emph{1-layer} i.e. homogeneous &    
    m_{1-layer}(z) &= m_{0}    & 
    \label{equa:m_hom}  \\
  
\emph{two-exponential} &
    m_{2-exp}(z) &=  m_{up} \cdot e^{-\frac{z-z_{up}}{\sigma_{up}}} + m_{0} + m_{low} \cdot e^{\frac{z-z_{low}}{\sigma_{low}}} 
    &  \label{equa:d-exp_d} = \ref{equa:d-exp}  \\

\emph{single-Gauss} &  
    m_{gauss}(z) &= m_{0} + m_{G} \cdot e^{\big(\frac{z-z_{G}}{\sigma_{G}}\big)^2} &
    \label{equa:gauss}   \\

\emph{single-exponential} &   
    m_{exp}(z) &= m_{up} \cdot e^{-\frac{z-z_{up}}{\sigma_{up}}} + m_{0} & 
    \label{equa:surf-exp}   \\  
\end{tabular*}
\caption{Analytical functions used to model the experimental asymmetry. $m_{0}$, is the magnetization in $\mu_B$ inside the layer and is rescaled to be equal to Fe bulk magnetic moment: 2.2 $\mu_B$. $m_{up}$, $\sigma_{up}$, $m_{low}$, and $\sigma_{low}$ correspond to the amplitude and width of the change of magnetization for the exponential profile at the upper and lower interfaces $z_{up}$ and $z_{low}$, and derived from the structural fit \cite{chardonnet_toward_2021}. $m_G$ and $\sigma_G$ are the amplitude and width of the Gaussian contribution in the single Gauss model, which has a free height of $z_G$. The magnetic fitting parameters are therefore, $m_{0}$, $m_{up}$, $\sigma_{up}$, $m_{low}$, $\sigma_{low}$, $m_G$, $z_G$ and $\sigma_G$.} \label{tab: modele_dynamique}
\end{ruledtabular}
\end{table*}

\begin{figure}[h!]
\includegraphics[width=18cm]{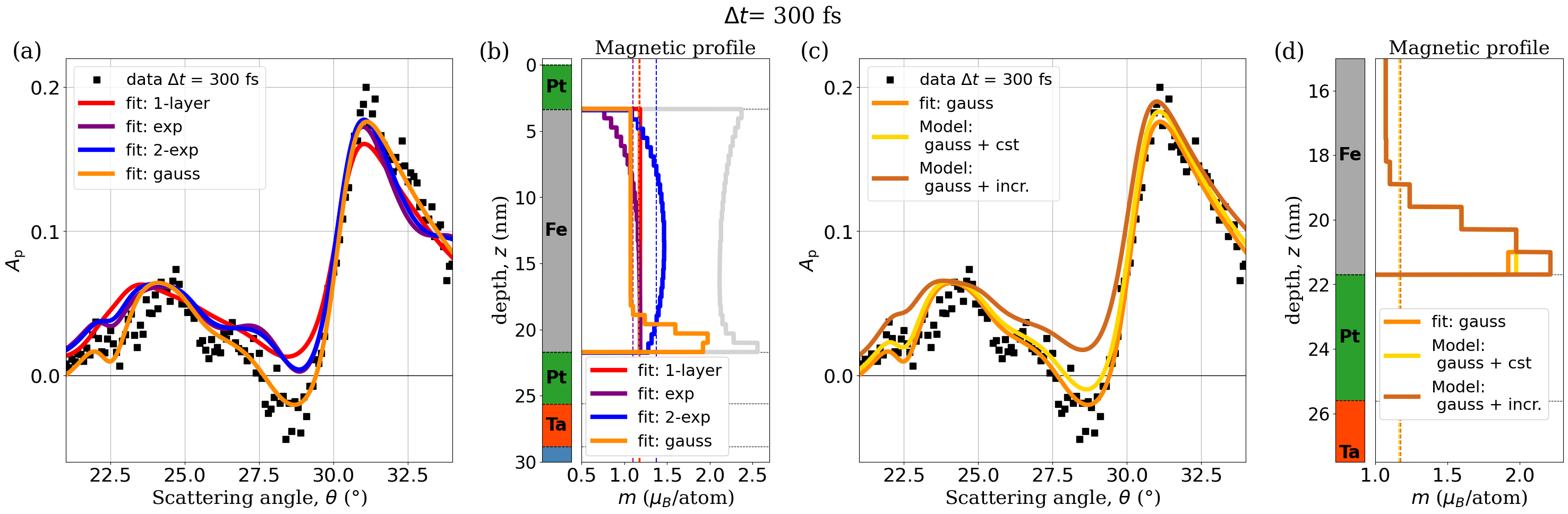}
\caption{\label{figS7} (a) and (c) Pumped asymmetry $A_\text{p}$ for a time delay $\Delta t =$ 300 fs (black squares), measured at FLASH at 701 eV with an energy resolution of 7 eV, with (a) the best fits (lines) obtained for the \emph{1-layer} (red), \emph{single-exponential} (purple), \emph{two-exponential} (blue) and \emph{single-Gauss} (orange) models and (c) the simulations corresponding to a constant (yellow line) and increased (brown line) magnetization in the last 7\r{A} close to the bottom Pt interface, as shown in the magnetic profiles displayed in (d). (b) Magnetic depth profiles corresponding to the fits reproducing the pumped asymmetry $A_\text{p}$ shown in (a). In light grey, the \emph{two-exponential} model corresponding to the umpumped magnetic profile is shown for comparison. (d) Magnetic depth profile around the bottom Pt interface, corresponding to the fit and model of the pumped asymmetry $A_\text{p}$ shown in (c).
}
\end{figure}

As an example, Fig. \ref{figS7}~a shows the experimental pumped asymmetry $A_\text{p}$ for a time delay $\Delta t =$ 300 fs (black squares), and the best fits for the four magnetic profiles of Table \ref{tab: modele_dynamique}. The corresponding magnetic profiles are shown in Fig. \ref{figS7}~b. For this time delay, the best agreement is clearly achieved with the Gaussian-type profile (orange line), which is the only one to show a negative asymmetry around $28\degree$.

As shown in fig. \ref{figS7}~b, the \emph{single-Gauss} model displays a depth-independent loss of magnetization over the first 15\,nm followed by a \emph{persistent magnetization band} centered at 2 nm from the bottom interface, with a final small magnetization decrease over the last 7\r{A} close to the bottom Pt layer. Note that the height of the Gaussian is a free parameter, and that we use a genetic algorithm to avoid being trapped in local minima during the fitting procedure, making us confident about this additional decrease in magnetization in the close vicinity of the lower Pt interface, that we also found for time delay $\Delta t$= 500\,fs and 1\,ps (Figs. \ref{figS12}~f and \ref{figS15}~g). To further investigate this decrease in the last 7\r{A}, from the \emph{single-Gauss} model, we changed the magnetic value of the last magnetic sub-layer to model a constant magnetization and an increased one in the last 7\r{A} close to the bottom Pt interface. Figures \ref{figS7}~c and d show the corresponding simulations and magnetic profiles close to the bottom interface. Fig. \ref{figS7}~c demonstrates that to reach the negative asymmetry data points around $28\degree$, one needs a magnetization decrease in the last 7\r{A} near the bottom Pt interface, which seems correctly described thanks to the \emph{single-Gauss} model. Note that in the transverse configuration used for the time-resolved experiment, we can't differentiate whether this loss of magnetization is only due to a reduction of the amplitude or to an out-of-plane rotation of the magnetization. This highlights the need to combine reflectivity geometry with sources providing sufficiently bright circularly polarized ultra-short soft X-rays.

\vspace{1cm}

\begin{figure}[h!]
	\begin{center}
		\begin{tabular}{c}
			  \includegraphics[height=6.6cm]{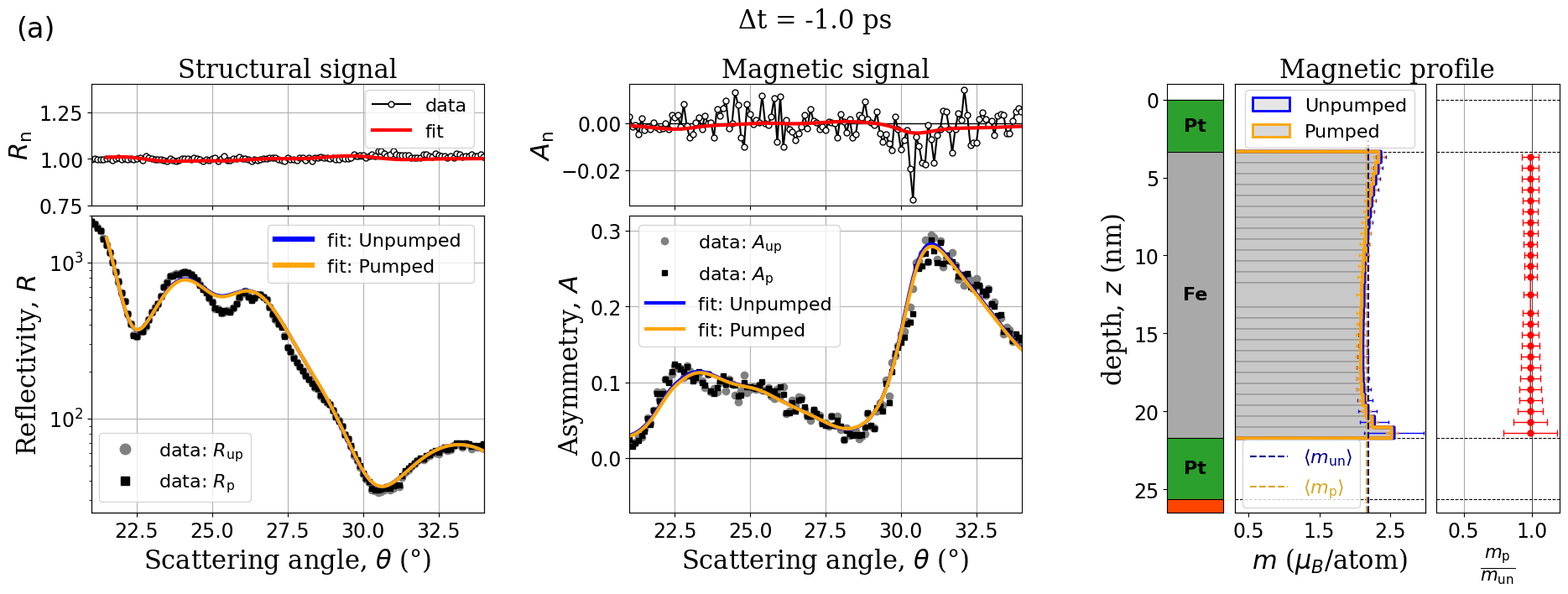} \\ 
		      \includegraphics[height=6.6cm]{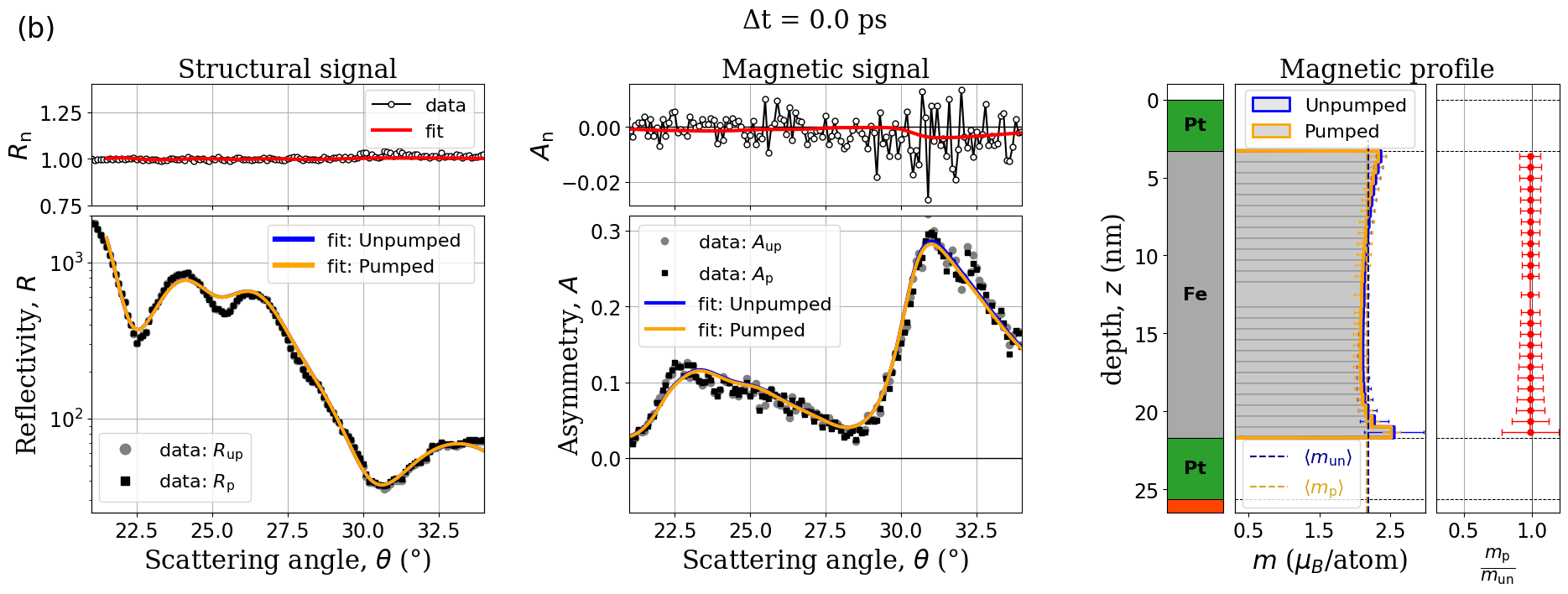} \\ 
			\includegraphics[height=6.6cm]{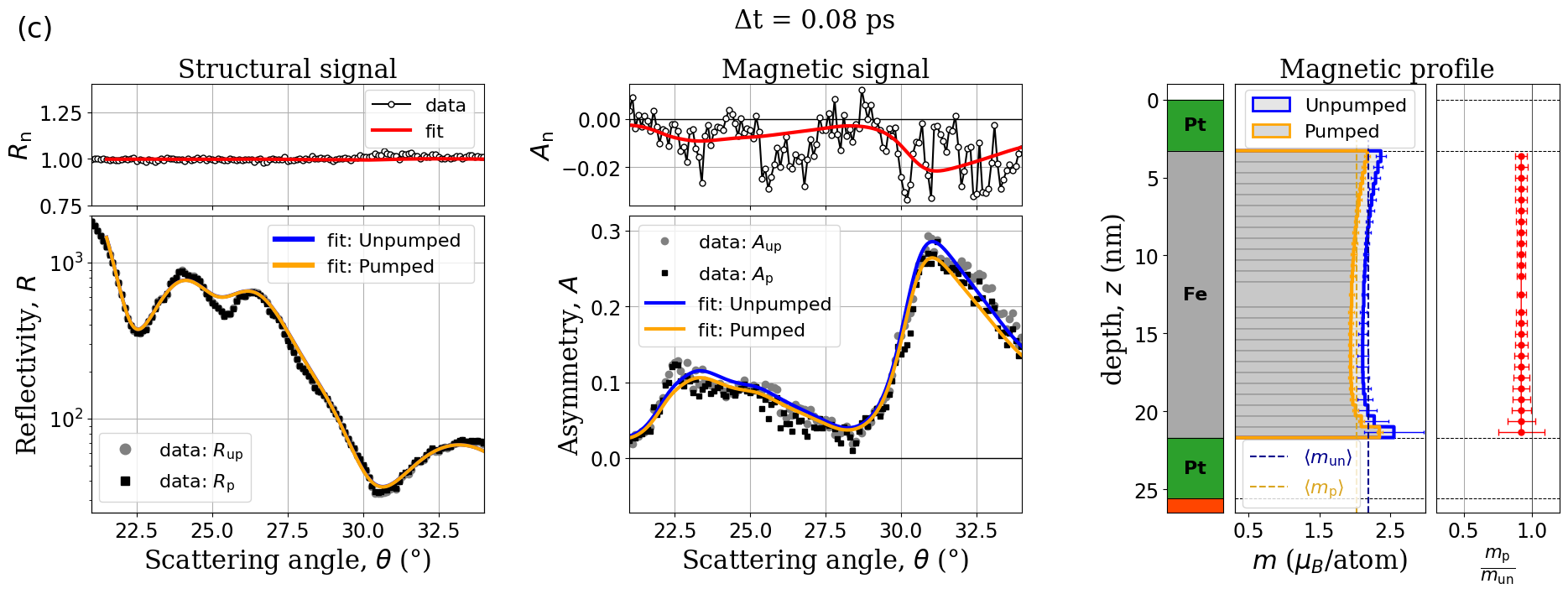}
		\end{tabular}
		\caption{Experimental data and quantitative analysis for (a) $\Delta t=-1$\,ps, (b) $\Delta t=0$\,ps, (c) $\Delta t=80$\,fs. The right panel shows the unpumped and pumped reflectivity $R_\text{up}$ (grey dots) and $R_\text{p}$ (black squares) as well as the normalized reflectivity $R_\text{n}=R_\text{p}/R_\text{up}$ (top panel) with their corresponding fits (line). The middle panel displays the unpumped and pumped asymmetry, $A_\text{up}$ and $A_\text{p}$ as well as the normalized asymmetry $A_\text{n}=A_\text{p} - A_\text{up}$ (top panel) with their corresponding best fits. The right panel shows the corresponding magnetic depth profile for pumped (orange line) and unpumped data (blue line) as well as  the normalized magnetization profile $m_\text{p}/m_\text{up}$ (red line).}
		\label{figS9}
	\end{center}
\end{figure}

\begin{figure}[h!]
	\begin{center}
		\begin{tabular}{c}
			  \includegraphics[height=6.6cm]{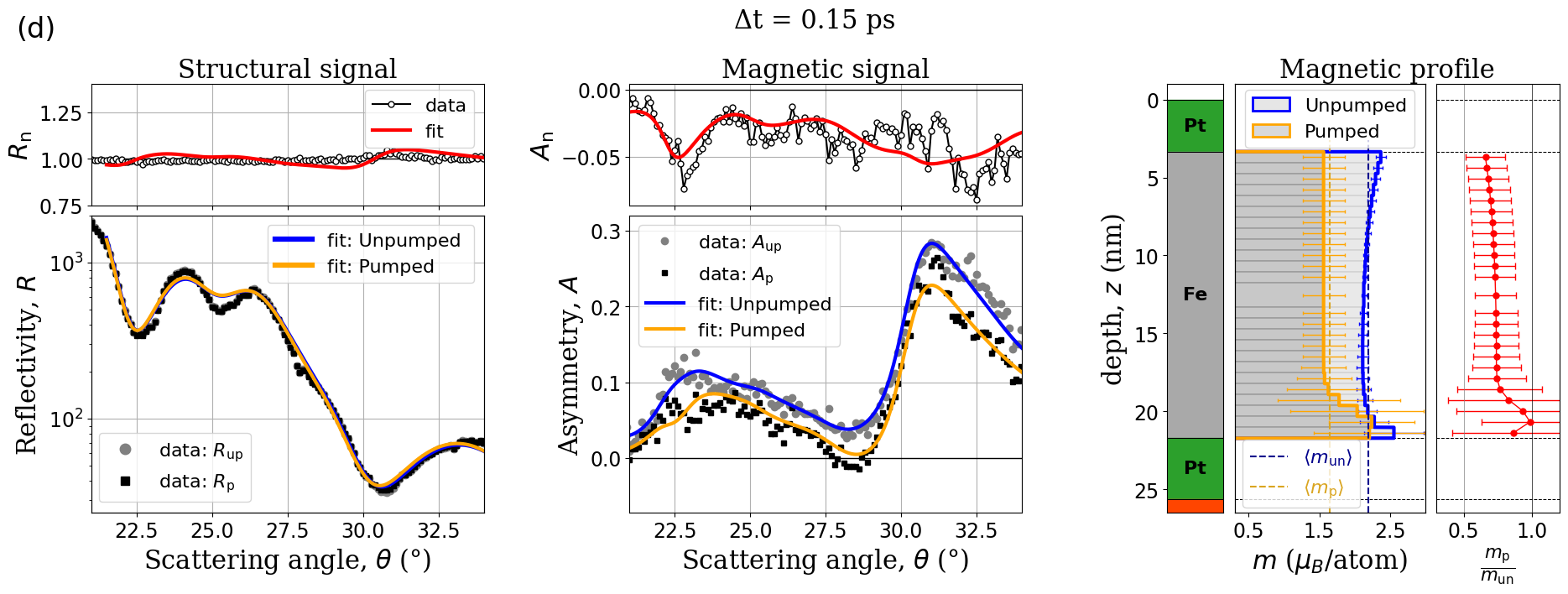} \\ 
		      \includegraphics[height=6.6cm]{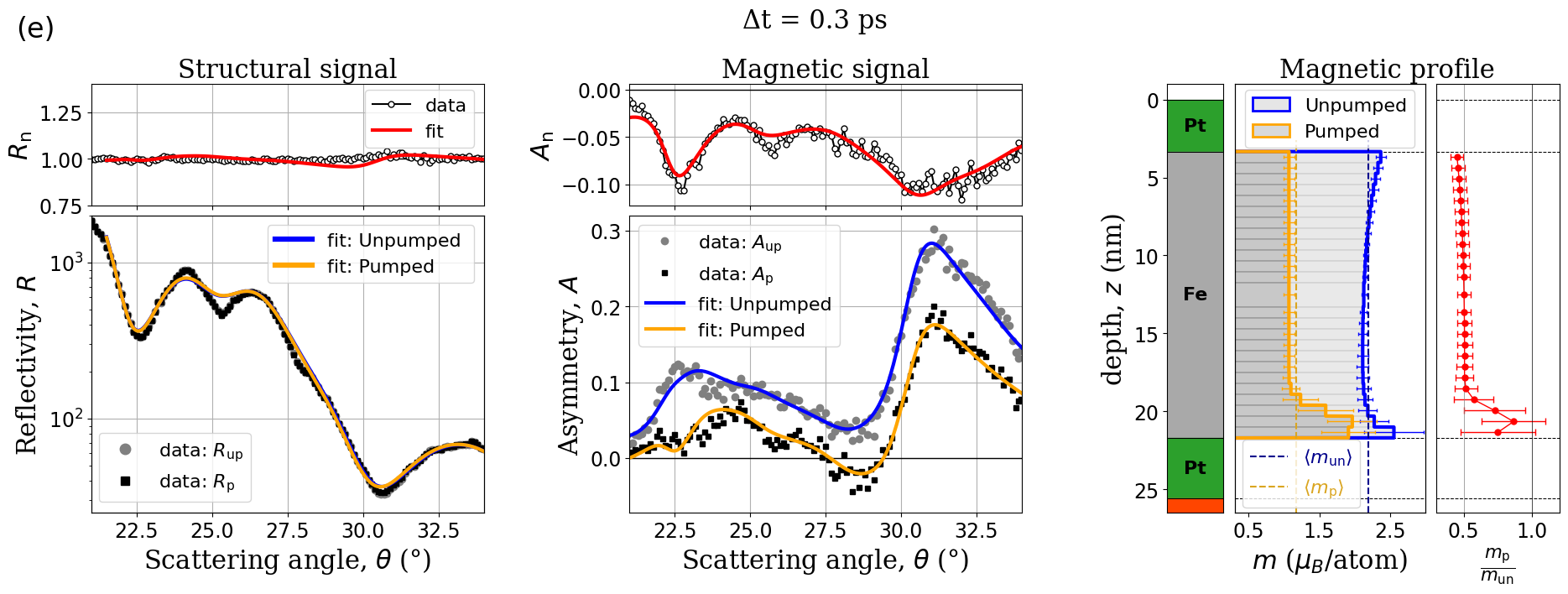} \\ 
			\includegraphics[height=6.6cm]{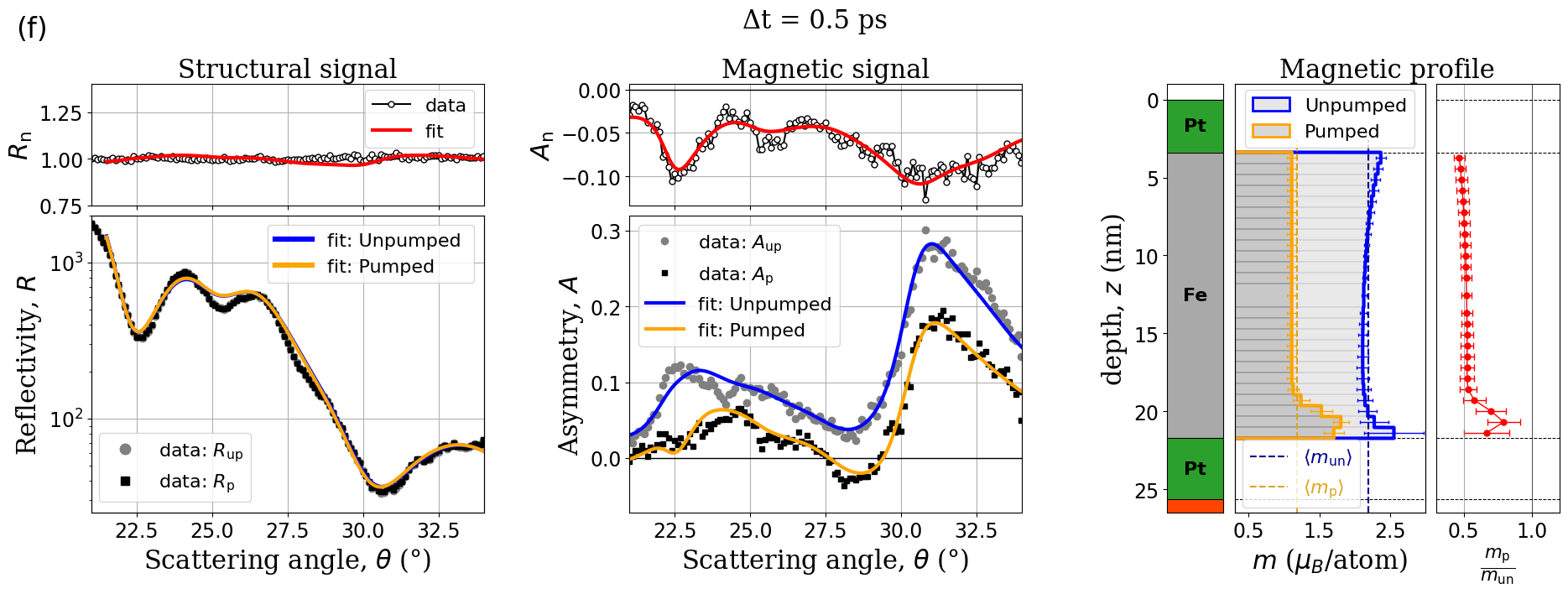}
		\end{tabular}
		\caption{Experimental data and quantitative analysis for (d) $\Delta t=150$\,fs, (e) $\Delta t=300$\,fs, (f) $\Delta t=500$\,fs. The right panel shows the unpumped and pumped reflectivity $R_\text{up}$ (grey dots) and $R_\text{p}$ (black squares) as well as the normalized reflectivity $R_\text{n}=R_\text{p}/R_\text{up}$ (top panel) with their corresponding fits (line). The middle panel displays the unpumped and pumped asymmetry, $A_\text{up}$ and $A_\text{p}$ as well as the normalized asymmetry $A_\text{n}=A_\text{p} - A_\text{up}$ (top panel) with their corresponding best fits. The right panel shows the corresponding magnetic depth profile for pumped (orange line) and unpumped data (blue line) as well as  the normalized magnetization profile $m_\text{p}/m_\text{up}$ (red line). }
		\label{figS12}
	\end{center}
\end{figure}

\begin{figure}[h!]
	\begin{center}
		\begin{tabular}{c}
			  \includegraphics[height=6.6cm]{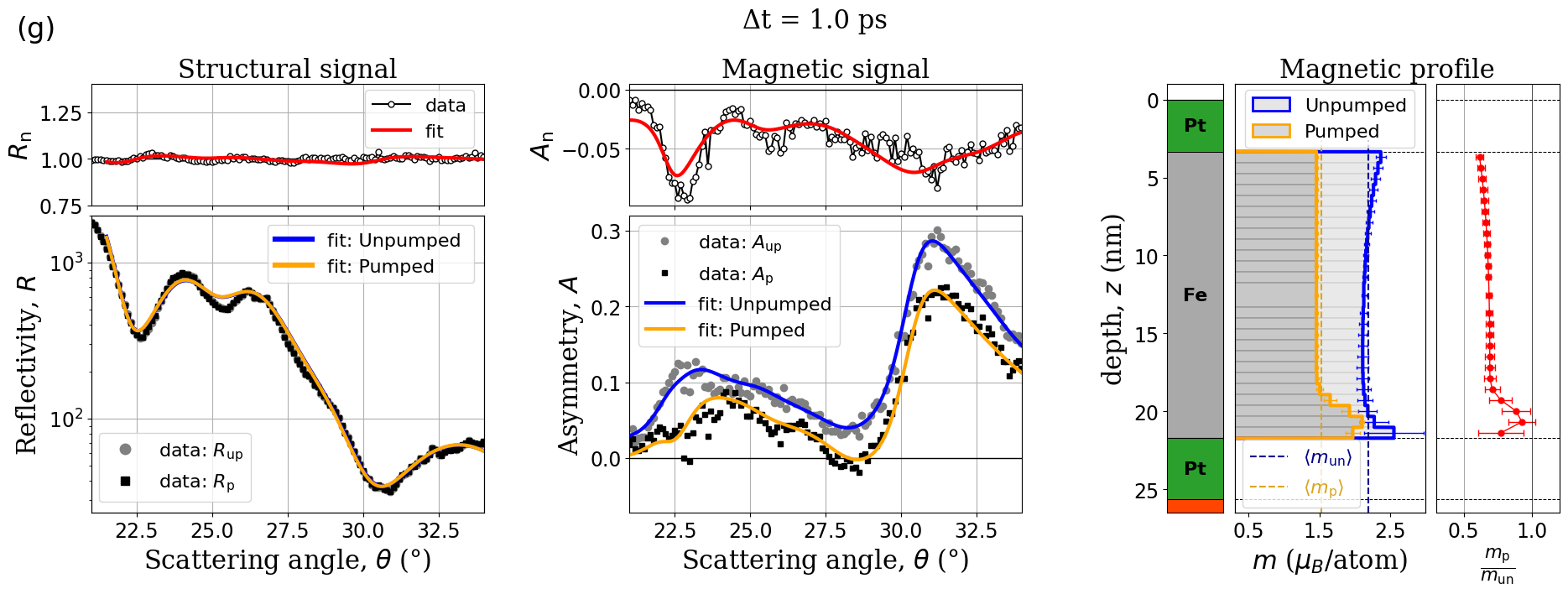} \\ 
		      \includegraphics[height=6.6cm]{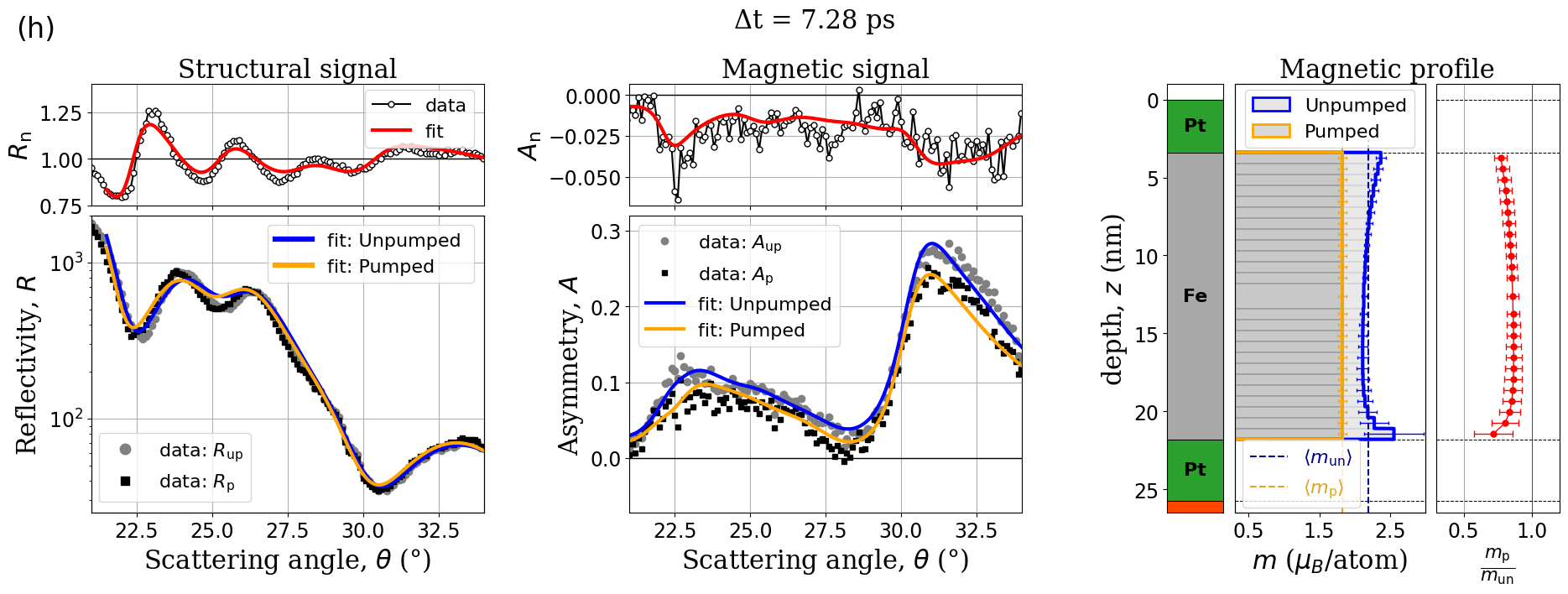} \\ 
			\includegraphics[height=6.6cm]{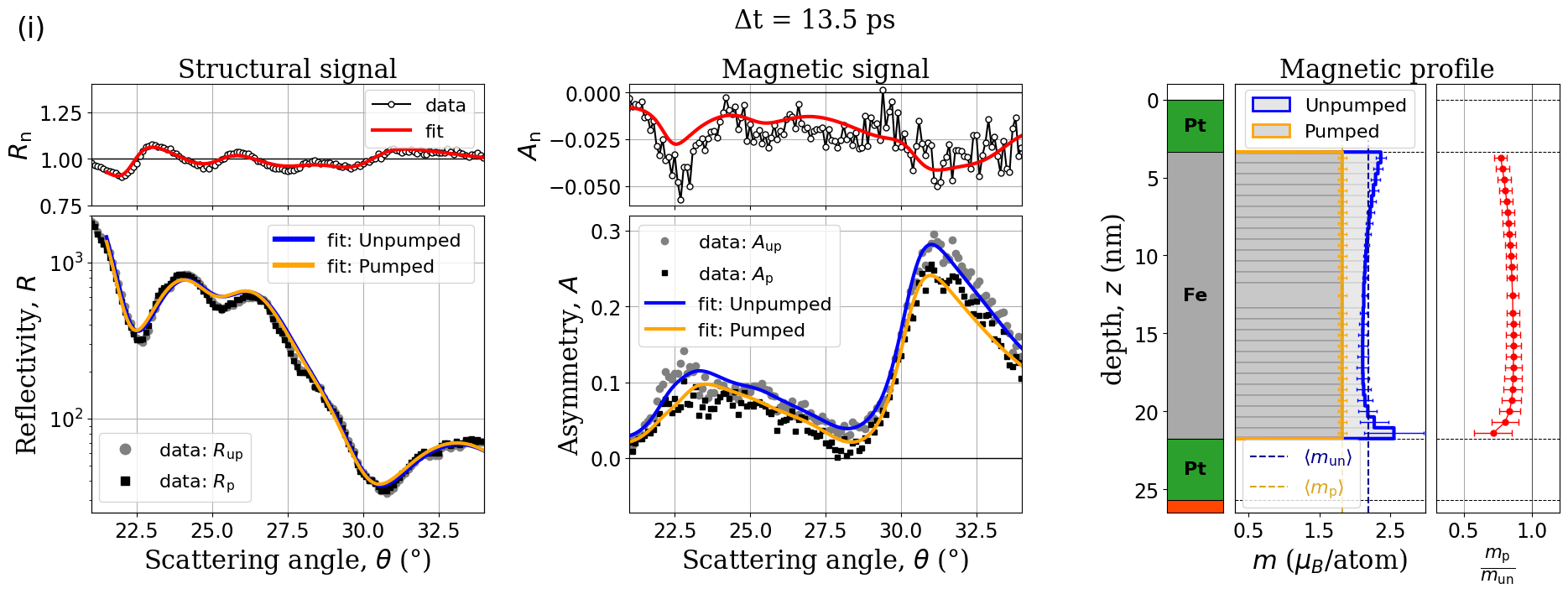}
		\end{tabular}
		\caption{Experimental data and quantitative analysis for (g) $\Delta t=1$\,ps, (h) $\Delta t=7.28$\,ps, (i) $\Delta t=13.5$\,ps. The right panel shows the unpumped and pumped reflectivity $R_\text{up}$ (grey dots) and $R_\text{p}$ (black squares) as well as the normalized reflectivity $R_\text{n}=R_\text{p}/R_\text{up}$ (top panel) with their corresponding fits (line). The middle panel displays the unpumped and pumped asymmetry, $A_\text{up}$ and $A_\text{p}$ as well as the normalized asymmetry $A_\text{n}=A_\text{p} - A_\text{up}$ (top panel) with their corresponding best fits. The right panel shows the corresponding magnetic depth profile for pumped (orange line) and unpumped data (blue line) as well as  the normalized magnetization profile $m_\text{p}/m_\text{up}$ (red line). }
		\label{figS15}
	\end{center}
\end{figure}

To summarize, for each time delay, we fitted $R_\text{up}$ and $A_\text{up}$ simultaneously, and $R_\text{p}$ and $A_\text{p}$ simultaneously for four different depth magnetic profiles: \emph{1-layer} (i.e. homogeneous), \emph{two-exponential}, \emph{single-exponential} and \emph{single-Gauss}, described by four analytic functions that are summarized in Table \ref{tab: modele_dynamique}. For each time delay, we selected the model that gave the best agreement between experimental and calculated reflectivities and asymmetries, as shown in Figs. \ref{figS9}~a to \ref{figS15}~i. These figures display the data and best-fit results for the reflectivity (left panel) and asymmetry (middle panel) as well as the corresponding magnetic depth profile (right panel) for pumped (black squares, orange line) and unpumped data (grey circles, blue line). 
We can see that for the unpumped asymmetry $A_\text{up}$, the \emph{two-exponential} fit is always in very good agreement, and within the error bar of the static \emph{two-exponential} function derived from our static study (equation \ref{equa: double exp soleil}, section \ref{sec:SOLEIL_fit}). For the pumped asymmetry $A_\text{p}$, the magnetization depth profile models that give the best fit depend on the time delay and are displayed in the right panel of \ref{figS9}~a to \ref{figS12}~i. The corresponding fit functions are displayed in table \ref{tab:pump-fit}. A summary of all those data and corresponding fits are shown in fig. 2 of the main article which displays the relative change (normalized signal) of the reflectivity and magnetic asymmetry: $R_\text{n} = R_\text{p} / R_\text{up}$ and $A_\text{n} = A_\text{p} - A_\text{up}$, also shown in the top panels of fig. \ref{figS9}~a to \ref{figS15}~i (open circles and red lines). Note that for the normalized asymmetry, $A_\text{n}$, we took the difference of pumped and unpumped data and not the ratio to avoid dividing by values close to 0 around $28\degree$ and get a better signal-to-noise ratio.

\begin{table*}[h!]
\begin{ruledtabular}
\begin{tabular*}{\textwidth}{
  @{\extracolsep{\fill}}
  l 
  l
  >{$\displaystyle}r<{\vphantom{\sum_{1}{N}}$} 
  @{\extracolsep{0pt}}
  >{$\displaystyle{}}l<{\vphantom{\sum_{1}{N}}$} 
  @{\extracolsep{\fill}}
  >{\refstepcounter{equation}(\theequation)}r
  @{}}
$\Delta t$ (\SI{}{\pico\second}) & $\langle m_\text{p}\rangle$ $(\mu_{B}/\text{at})$  & \multicolumn{2}{l}{Best fit function $m(z)$ $(\mu_{B}/\text{at})$ } & \multicolumn{1}{c}{} \\
\midrule
-1.00 & 2.17 & m_{2-exp}(z) &
            = 0.30(\pm0.05) \cdot e^{-\frac{z-3.2}{6.0(\pm1.0)}} + 2.07(\pm0.03) + 0.70(\pm0.10) \cdot e^{\frac{z-21.6}{1.1(\pm0.5)}}  
            &  = \ref{equa:d-exp}  \\
0.00 & 2.17 & m_{2-exp}(z) &
            = 0.30(\pm0.05) \cdot e^{-\frac{z-3.2}{6.0(\pm1.0)}} + 2.07(\pm0.03) + 0.70(\pm0.10) \cdot e^{\frac{z-21.6}{1.1(\pm0.5)}}  
            &  = \ref{equa:d-exp}  \\
0.08 & 2.02 & m_{2-exp}(z) &
            = 0.28(\pm0.05) \cdot e^{-\frac{z-3.3}{6.0(\pm1.0)}} + 1.9(\pm0.03) + 0.64(\pm0.10) \cdot e^{\frac{z-21.7}{1.1(\pm0.5)}}  
            &  \label{equa:t=0.08}$\approx 0.92 \times$ \ref{equa:d-exp}  \\
0.15 & 1.65 & m_{gauss}(z) &
            = 1.57(\pm0.07) + 0.71(\pm0.02) \cdot e^{\big(\frac{z-20.9(\pm0.3)}{1.6(\pm0.2)}\big)^2}
            & \label{equa:t=0.15}   \\
0.3  & 1.17 & m_{gauss}(z) &
            = 1.07(\pm0.09) + 0.94(\pm0.20) \cdot e^{\big(\frac{z-20.9(\pm0.2)}{1.3(\pm0.1)}\big)^2}
            & \label{equa:t=0.3}   \\
0.5  & 1.18 & m_{gauss}(z) &
            = 1.11(\pm0.06) + 0.71(\pm0.06) \cdot e^{\big(\frac{z-20.9(\pm0.1)}{1.2(\pm0.1)}\big)^2}
            & \label{equa:t=0.5}   \\
1.00  & 1.53 & m_{gauss}(z) &
            = 1.46(\pm0.02) + 0.65(\pm0.01) \cdot e^{\big(\frac{z-20.8(\pm0.2)}{1.3(\pm0.1)}\big)^2}
            & \label{equa:t=1}   \\
7.28 & 1.83 & m_{1-layer}(z) &= 1.83(\pm0.03)    &      \label{equa:t=7.28}  \\

13.50 & 1.82 & m_{1-layer}(z) &= 1.82(\pm0.03)   &      \label{equa:t=13.5}  \\

\end{tabular*}
\caption{ Summary of the best fit functions obtained for each time delay $\Delta t$ that are displayed in Figs. \ref{figS9}, \ref{figS12}, \ref{figS15} and 2 of the article. The average pump values $\langle m_\text{p}\rangle$ used for the figure 4~a of the article are also displayed.} \label{tab:pump-fit}
\end{ruledtabular}
\end{table*}

\begin{figure}[h!]
\includegraphics[width=15cm]{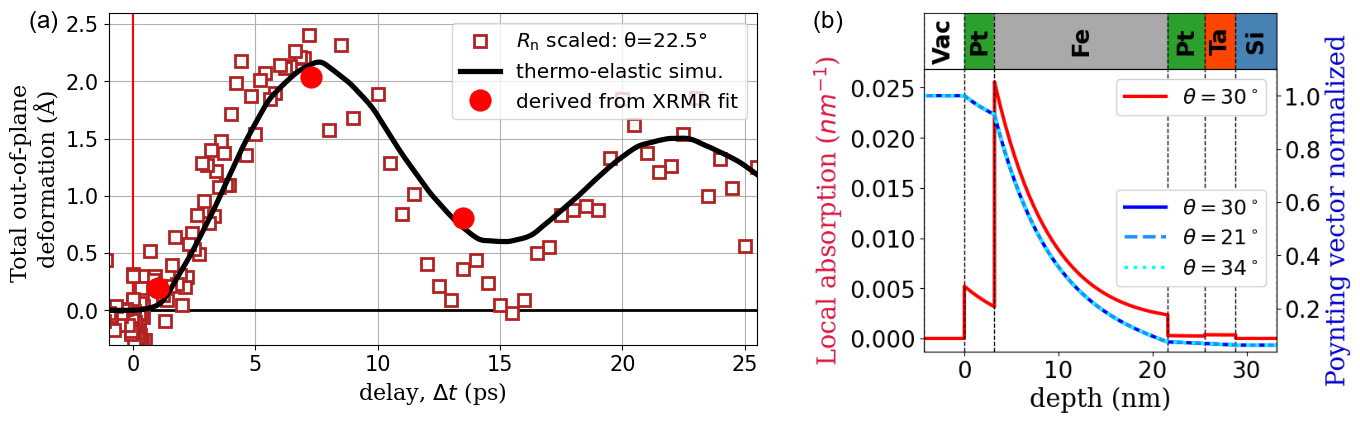}
\caption{\label{figS18} (a) Laser-induced total out-of-plane deformation as a function of time delay, extracted from the $R_n$ fits of Fig. \ref{figS9}~a to \ref{figS12}~i, in red dots. The black line shows the total deformation calculated with the udkm1Dsim toolbox. The dark red squares represent a time delay scan of the normalized reflectivity, $R_\text{n}(\Delta t)$, performed at $\theta=22.5\degree$ and scaled to match the fitted value (red dots). (b) IR absorption profile: local absorption in red (left axis) and normalized Poynting vector (photon flux) in blue (right axis) as a function of the sample depth. The reflection angles are indicated in the legend. }
\end{figure}

\clearpage
The corresponding relative thickness change, i.e., the total out-of-plane deformation, extracted from those fits, for each layer and each delay, is shown in Fig. 2 b of the article. The average out-of-plane deformation over the entire thin film thickness, which is in very good agreement with the thermo-elastic simulation, as shown in Fig. 3 of the article and in Fig. \ref{figS18}~a can also be compared to a time delay scan of the normalized reflectivity $R_\text{n}(\Delta t)$ performed at a specific reflectivity angle of $\theta=22.5\degree$ (red squares in Fig. \ref{figS18}~a). While at a fixed angle, $R_\text{n}(\Delta t)$ is not exactly proportional to the thin film thickness, scaling those data to the derived out-of-plane deformation strengthens the good agreement between the experimental data and the thermo-elastic simulation in terms of oscillations periodicity (Fig. \ref{figS18}~a).

\begin{figure}[h!]
   \includegraphics[width=0.75\linewidth]{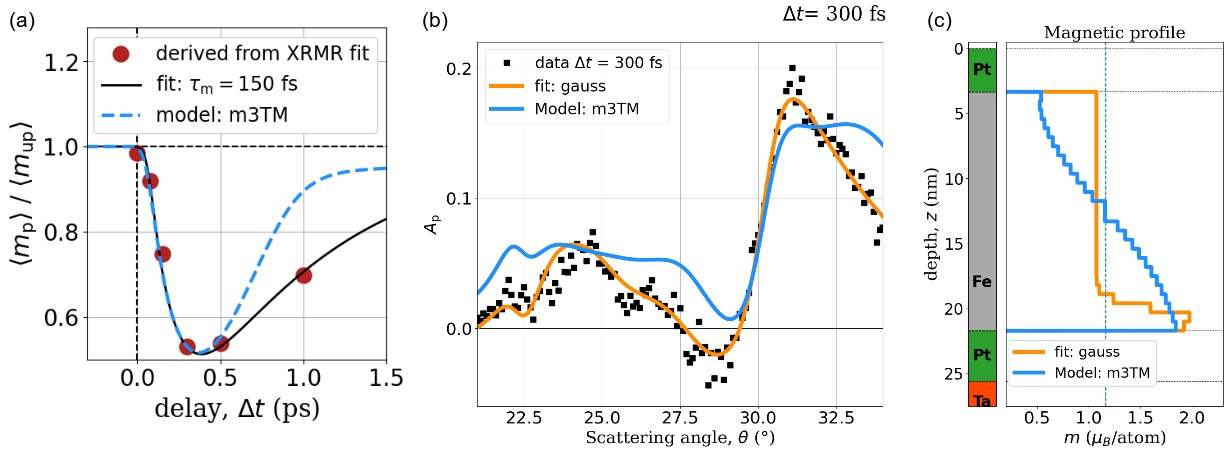}
   \caption{(a) Time evolution of the average relative magnetization of iron $\langle m_\text{p} \rangle/ \langle m_\text{up}\rangle$: the magnetization measured by tr-XRMR is represented by red dots while the black line is the result of a fit (equation \ref{equa:d-exp_d}) and the dotted blue line is the result of the m3TM simulation. (b) Pumped asymmetry $A_\text{p}$ for a time delay $\Delta t =$ 300 fs (black squares), measured at FLASH at 701 eV with an energy resolution of 7 eV, with the best fit obtained for the \emph{single-Gauss} model (orange line) and the simulations corresponding to the m3TM magnetic profile derived from udkm1Dsim caluculations (blue line). (c) Magnetic depth profile corresponding to the fit and model of the pumped asymmetry $A_\text{p}$ shown in (b) with the same color code.}
   \label{figS21}
\end{figure}

The average ultrafast demagnetization values of the Fe layer, $\langle m_\text{p} \rangle$, can be obtained by averaging the depth profiles. They are shown as dashed lines in the right panel of Figs. \ref{figS9}~a to \ref{figS12}~i, and are displayed in Table \ref{tab:pump-fit} ($\langle m_\text{up}\rangle=2.2\,\mu_{B}/\text{at}$). When $\langle m_\text{p} \rangle/ \langle m_\text{up}\rangle$ is plotted as a function of the delay, we retrieved a demagnetization curve (Fig. 4~a of the article and \ref{figS21}~a) that we fitted by a double exponential function (equation \ref{demag fit}) convoluted with a Gaussian function of 70\,fs full width half maximum. The result of the fit (black line in Fig. 4~a of the article and Fig. \ref{figS21}~a) gives a demagnetization amplitude $A=0.42$, a recovery time $\tau_\text{r}=900$\,fs and a demagnetization time $\tau_\text{m}=150$\,fs. 
\begin{align}
    \frac{\langle m_\text{p}(t>0) \rangle}{\langle m_\text{up}\rangle} = 1-A(\text{e}^{-t/\tau_\text{r}}-\text{e}^{-t/\tau_\text{m}})
    \label{demag fit}
\end{align}

Finally, the normalized magnetization profiles, showing $m_\text{p}/m_\text{up}$ as a function of the depth, and corresponding to the ratio of the orange and blue profiles displayed in the right panel of Figs. \ref{figS9}~a to \ref{figS12}~i, are shown in the right insert (red points). We interpolated these normalized magnetization profiles to obtain the $m_\text{p}(z,\Delta t)/m_\text{up}(z)$ 2D map shown in Fig. 4~b of the article.

\section{\label{sec:Optic_exc_prof}Optical excitation profile}
The optical excitation profile was calculated for an 800\,nm infrared (IR) pump with the refractive index of Table \ref{tab : paramètre physique M3TM}, using the transfer matrix method package (\textit{tmm}, \cite{byrnes_multilayer_2020}). Figure \ref{figS18}~b shows the local absorption in red and the normalized Poynting vector in blue for a reflection angle $\theta = 30\degree$. Note that we took care of the angle-dependent footprint and absorption by modulating the intensity of the incoming infrared pulses as explained in our previous article \cite{chardonnet_toward_2021}. We also checked that the absorption profile doesn't change for different reflection angles (blue dotted lines in Fig. \ref{figS18}~b).

\section{\label{sec:udkm1Dsim}m3TM and udkm1Dsim simulation details}

\begin{figure}[h!]
   \includegraphics[width=0.75\linewidth]{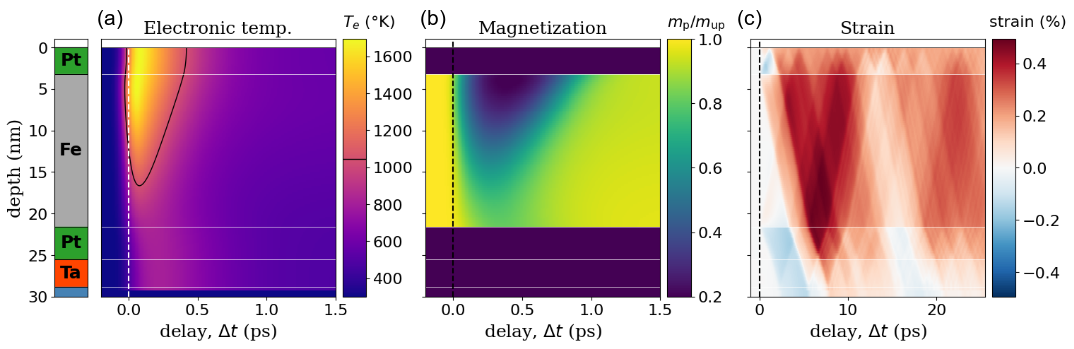}
   \caption{(a) Evolution of the electronic temperature $T_{e}$ as a function of sample depth and time delay, where the black contour highlights the Curie temperature of Fe. (b) Evolution of the normalized magnetization, $m_\text{p}/m_\text{up}$, as a function of sample depth and time delay (same as Fig. 4~c of the article). (c) Evolution of the optically induced strain as a function of sample depth and time delay.}
   \label{figS20}
\end{figure}

We used the \textbf{udkm1Dsim} toolbox \cite{schick_udkm1dsim_2021, schick_udkm1dsim_2024}, and the microscopic 3 temperature model (m3TM) \cite{koopmans_explaining_2010}, to calculate the temporal evolution of the electronic temperature (Fig. \ref{figS20}~a), magnetization (Fig. \ref{figS20}~b, same as fig 4~c in the main article) and strain (Fig. \ref{figS20}~c) as a function of depth resulting from laser-excitation of our multilayer. Contrary to the classical 3-temperature model \cite{beaurepaire1996ultrafast}, the m3TM describes the spin system not directly through the spin temperature but through a magnetization differential equation. More information on how this m3TM is implemented in the \textbf{udkm1Dsim} toolbox can be found in the \textbf{udkm1Dsim} documentation \cite{schick_udkm1dsim_m3tm}.

According to this simulation, following the sudden rise in temperature in the upper part of the iron layer (Fig. \ref{figS20}~a), stresses are generated in a very short time ($\Delta t < \SI{0.5}{\pico\second}$, \ref{figS20}~c). These stresses relax coherently, generating a shock wave that propagates through the sample at the speed of sound. From this 2D map we can calculate the total deformation of the Fe layer and extract the time traces plotted in Fig. \ref{figS18}~a and Fig. 3 of the article (black line). 
While the structural dynamics start immediately at the interface (Fig. \ref{figS20}~c), we can't access the transient strain depth profile in reflectivity, but only the total thickness changes over one elemental layer.

For the simulations shown in this work, we used the density and thicknesses retrieved in our analysis (table \ref{tab:struct}) and nominal values from the literature. Table \ref{tab : paramètre physique M3TM} summarizes all physical quantities and parameters needed to perform the \textbf{udkm1Dsim} simulations. Note that we multiplied the static thermal expansion coefficients by 2 to account for the Poisson stresses, as shown in Mattern \textit{et al.}. In our work, we did not vary or fit any parameters, which would probably improve the agreement with our experimental result, but is beyond the scope of this article. As shown in Fig. \ref{figS21}~a, during the first 500\,fs, the agreement between the normalized average moment derived from our quantitative analysis (red points) and that calculated with the m3TM (dashed blue line) is very good. However, the asymmetry calculated from the m3TM magnetization profile at a time delay of $\Delta t = 300$\,fs (blue line in Fig. \ref{figS21}~b and c) does not reproduce our experimental data at all. This highlights the need for depth-sensitive methods to validate the use of microscopic models.

\begin{table}[h!]
\begin{center}

\begin{tabular}{p{6cm}|p{2.6cm}|p{2.6cm}|p{2.6cm}|p{2.6cm}}
\multicolumn{5}{c}{Experimental parameters} \\ \hline \hline
\multicolumn{2}{l|}{Reflection angle $\theta$ (\SI{}{\degree})}& \multicolumn{3}{c}{30}
\\
\multicolumn{2}{l|}{Incident fluence $F_{inc}$ (\SI{}{\milli\joule\per\centi\meter\squared})} & \multicolumn{3}{c}{23.5} \\
\multicolumn{2}{l|}{pump wavelength (\SI{}{\nano\meter})}& \multicolumn{3}{c}{800}
\\
\multicolumn{2}{l|}{pulse width (\SI{}{\pico\second})} & \multicolumn{3}{c}{0.1}
\\ \hline \hline
\multicolumn{5}{c}{} \\
\multicolumn{5}{c}{Sample parameters} \\ \hline
 & Pt & Fe & Ta & Si\\ \hline \hline
density (\SI{}{\kilo\gram\per\meter\cubed}) from XRMR fit, tab. \ref{tab:struct} & 18923  & 6701  & 14114 & 2331 
\\ \hline
thickness (\SI{}{\nano\meter}) & 3.2 | 3.9 & 18.4 & 3.3 & 
\\ \hline
roughness (\SI{}{\nano\meter}) & 0.9 | 0.3 & 0.8 & 0.5 & 0.3
\\ \hline 
refractive index $n$ \cite{refractiveindex} & $0.57617 + 8.077 6\cdot i$ & $3.6794 + 10.413 \cdot i$  & $0.99181 + 7.2934 \cdot i$  & $3.6690 + 0.0052655 \cdot i $ 
\\ \hline
specific heat(\SI{}{\joule\per\meter\cubed\per\kelvin}) : & & & &
\\
$\bullet$ electrons $C_{e}$ & $748.1 \cdot T_{e}$ \cite{lin_electron-phonon_2008, hohlfeld_electron_2000, pudell_heat_2020} & $704.57 \cdot T_{e}$ \cite{gao_effects_1998} & $380 \cdot T_{e}$ \cite{pudell_heat_2020} & $150 \cdot T_{e}$ \cite{schick_udkm1dsim_2024}
\\
$\bullet$ phonons $C_{ph}$ & $2.8 \cdot 10^6$ \cite{schick_udkm1dsim_2024, pudell_heat_2020} & $3.49 \cdot 10^6$ \cite{gao_effects_1998} & $2.33 \cdot 10^6$ \cite{pudell_heat_2020} & $1.6 \cdot 10^6$ \cite{schick_udkm1dsim_2024}
\\ \hline
Thermal conductivity (\SI{}{\watt\per\meter\per\kelvin}): & & & &
\\
$\bullet$ electrons $\kappa_{e}$ & $66 \cdot T_{e}/T_{ph}$ \cite{pudell_heat_2020} & $65 \cdot T_{e}/T_{ph}$ \cite{backlund_experimental_1961} & $52 \cdot T_{e}/T_{ph}$ \cite{pudell_heat_2020} & 0 \cite{koroleva2017determination}
\\
$\bullet$ phonons $\kappa_{ph}$ & 5 \cite{pudell_heat_2020} & 7 \cite{backlund_experimental_1961} & 5 \cite{pudell_heat_2020} & $158500 \cdot T^{-1.23}$ \cite{glassbrenner_thermal_1964, van_driel_kinetics_1987}
\\ \hline
electron/phonon coupling $g_{ep}$ (\SI{}{\peta\watt\per\meter\per\kelvin}) & 400  \cite{lin_electron-phonon_2008, hohlfeld_electron_2000, medvedev_electron-phonon_2020} & 2500 \cite{henighan_generation_2016, medvedev_electron-phonon_2020} & 100 \cite{pudell_heat_2020, medvedev_electron-phonon_2020} & 0 \cite{medvedev_electron-phonon_2020}
\\ \hline
material dependant scale factor $R$ (\SI{}{\per\pico\second}) & 0 & 25.3 \cite{wieczorek_separation_2015} & 0 & 0
\\ \hline
Curie temperature $T_{C}$(\SI{}{\kelvin}) & 0 & 1043 \cite{noakes_initial_1966} & 0 & 0
\\ \hline
linear thermal expansion coefficient $\alpha$ ($\SI{2e-6}{\per\kelvin}$)& 9 \cite{hahn_thermal_1972, nix_thermal_1942} & 12 \cite{kozlovskii_linear_2019} & 6.5 \cite{nix_thermal_1942} & 2.6 \cite{white_thermal_1973}
\\ \hline
speed of sound $v_{s}$ (\SI{}{\nano\meter\per\pico\second}) \cite{WebElements}& 2.680  & 4.910  & 3.400  & 8.433
\\ \hline
\end{tabular}
	\caption{\small{Summary of all the physical constants and parameters required for simulating the m3TM model. The 2 values for the thickness and roughness of the platinum correspond to the top and bottom layer values, respectively.}}
	\label{tab : paramètre physique M3TM}
\end{center}
\end{table}


\bibliography{biblio}  

\end{document}